\documentclass[prc,superscriptaddress,unsortedaddress,twocolumn,showpacs,preprintnumbers,amsmath,amssymb,floatfix]{revtex4}
\usepackage[dvips]{graphicx}
\usepackage{amsmath}
\usepackage{amssymb}
\usepackage{mathrsfs}
\usepackage{multirow}
\usepackage{bm}
\usepackage{wrapft}
\usepackage{color}

\def\fun#1#2{\lower3.6pt\vbox{\baselineskip0pt\lineskip.9pt
\ialign{$\mathsurround=0pt#1\hfil##\hfil$\crcr#2\crcr\sim\crcr}}}

\def\a{\alpha}

\newcommand{\beq}{\begin{equation}}
\newcommand{\eeq}{\end{equation}}
\newcommand{\bea}{\begin{eqnarray}}
\newcommand{\eea}{\end{eqnarray}}

\newcommand{\bfi}[1]{\mbox{\boldmath $#1$}}

\newcommand{\vK}{{\bfi K}}

\newcommand{\vs}{{\bfi s}}

\newcommand{\vrr}{{\bfi r}}
\newcommand{\vR}{{\bfi R}}

\newcommand{\vX}{{\bfi X}}

\begin{document}


\title{Deformation of Ne isotopes in the island-of-inversion region}

\author{Takenori Sumi}
\affiliation{Department of Physics, Kyushu University, Fukuoka 812-8581, Japan}

\author{Kosho Minomo}
\affiliation{Department of Physics, Kyushu University, Fukuoka 812-8581, Japan}

\author{Shingo Tagami}
\affiliation{Department of Physics, Kyushu University, Fukuoka 812-8581, Japan}

\author{Masaaki Kimura}
\affiliation{Creative Research Institution (CRIS), Hokkaido University, Sapporo 001-0021, Japan}

\author{Takuma Matsumoto}
\affiliation{Department of Physics, Kyushu University, Fukuoka 812-8581, Japan}

\author{Kazuyuki Ogata}
\affiliation{Research Center of Nuclear Physics (RCNP), Osaka University, Ibaraki 567-0047, Japan}

\author{Yoshifumi R. Shimizu}
\affiliation{Department of Physics, Kyushu University, Fukuoka 812-8581, Japan}

\author{Masanobu Yahiro}
\affiliation{Department of Physics, Kyushu University, Fukuoka 812-8581, Japan}

\date{\today}

\begin{abstract}
The deformation of Ne isotopes in the island-of-inversion region
is determined by the double-folding model 
with the Melbourne $g$-matrix and the density calculated
by the antisymmetrized molecular dynamics (AMD).
The double-folding model reproduces, with no adjustable parameter,
the measured reaction cross sections for 
the scattering of $^{28-32}$Ne from $^{12}$C at 240MeV/nucleon.
The quadrupole deformation thus determined is around 0.4
in the island-of-inversion region and $^{31}$Ne is a halo nuclei
with large deformation.
We propose the Woods-Saxon model
with a suitably chosen parameterization set and the deformation given
by the AMD calculation
as a convenient way of simulating the density calculated directly by the AMD.
The deformed Woods-Saxon model provides the density
with the proper asymptotic form.
The pairing effect is investigated, and the importance of
the angular momentum projection for obtaining the large deformation
in the island-of-inversion region is pointed out.
\end{abstract}


\maketitle
\section{Introduction}

Exploring the so-called ``Island of inversion'' is one of the most
important current subjects in nuclear physics.
The term ``Island of inversion (IOI)'' was first introduced
by Warburton~\cite{Warburton} to
the region of unstable nuclei from $^{30}$Ne to $^{34}$Mg.
In the region, the low excitation energies and the large $B(E2)$ values
of the first excited states suggest
strong deformations~\cite{Mot95,Caurier,Utsuno,Iwas01,Yana03}.
This indicates that the $N=20$ magic number is no longer valid.
This novel quantum property has triggered
extensive experimental and theoretical studies on the IOI region.

Another important progress of research on unstable nuclei is the discovery of
the halo structure~\cite{Tanihata,Jensen,Jonson}.
Very recently, the interaction cross section $\sigma_{\rm I}$ was measured
by Takechi {\it et al.}~\cite{Takechi} for the scattering of 
$^{28-32}$Ne at 240~MeV/nucleon and 
it was shown that $\sigma_{\rm I}$ is quite large
for $^{31}$Ne. A halo structure of $^{31}$Ne was reported by the
experiment on the one-neutron removal reaction~\cite{Nakamura}.
This is the heaviest halo nucleus in the present
stage suggested experimentally. The nucleus resides in the IOI region.
The reaction cross section $\sigma_{\rm R}$ (or $\sigma_{\rm I}$)
and the nucleon-removal cross section with radioactive
beams are thus important experimental tools for exploring unstable
nuclei~\cite{Tanihata,Jensen,Jonson,Takechi,Nakamura,Gade}. 
For the scattering of unstable nuclei at intermediate energies, 
$\sigma_{\rm I}$ agrees with $\sigma_{\rm R}$ exactly or nearly, 
since projectile excitations to its discrete excited states 
do not exist or small even if they exist. 
This is discussed in this paper.

A useful theoretical tool of analyzing $\sigma_{\rm R}$ is
the microscopic optical potential constructed
by the double-folding model (DFM) with
the $g$-matrix effective nucleon-nucleon (NN)
interaction~\cite{M3Y,JLM,Brieva-Rook,Satchler-1979,Satchler,CEG,
Rikus-von Geramb,Amos,CEG07},
when the projectile breakup is weak.
For the nucleon-nucleus scattering,
the single-folding model with the Melbourne $g$-matrix
well reproduces the data on $\sigma_{\rm R}$ and
the elastic-scattering cross section
systematically~\cite{Amos}.
For $^{31}$Ne scattering from
$^{12}$C at 240~MeV/nucleon, the breakup cross section is
about 1\% of $\sigma_{\rm R}$~\cite{ERT}.
The DFM is hence reliable also for other projectiles $^{20-32}$Ne, 
since $^{31}$Ne is a most-weakly bound system among them.

In the DFM, the microscopic optical potential is constructed by folding
the $g$-matrix with projectile and target densities.
The density profile changes, if it is deformed.
The elongation makes the surface diffuseness and
the root-mean-square (RMS) radius effectively larger and
eventually enhances $\sigma_{\rm R}$.
The amount of deformation is thus important.
Nuclei in the IOI region are spherical
or only weakly deformed in the Skyrme and/or Gogny HF (HFB) calculations;
see, e.g., Refs.~\cite{TFH97,RER02}.
It is even pointed out that the observed large $B(E2;2^+\rightarrow 0^+)$
values can be understood as a large amplitude vibration around
the spherical shape~\cite{YG04}.
In such a situation, the additional correlations by
the angular momentum projection (AMP) often leads to possible deformed shapes;
see Ref.~\cite{RER03} for Ne isotopes.

Recently a systematic analysis was made by
the antisymmetrized molecular dynamics (AMD)
with the Gogny D1S interaction
for both even and odd $N$ nuclei in the IOI region~\cite{Kimura,Kimura1}.
The AMP-AMD calculations, i.e. the AMD calculation with the AMP performed,
yields rather large deformations.
This is consistent with the AMP-HFB calculations~\cite{RER02,RER03}.
A consistent picture of even and odd isotopes has been obtained by
the AMP-AMD approach,
where the $n$-particle $m$-hole excitations of the Nilsson orbits
play important roles to determine deformed configurations.
Although it is difficult to distinguish the dynamic shape-fluctuation
and static deformation in these light mass nuclei,
one may use the deformed shape suggested by the AMD calculation
to see its effect on $\sigma_{\rm R}$.
Very recently the Woods-Saxon mean-field model with the deformation obtained by
the AMP-AMD calculation was applied for $^{28-32}$Ne and the DFM 
with the density of the mean-field model was successful in
reproducing the data on $\sigma_{\rm R}$
in virtue of the deformation~\cite{Minomo-DWS}.

In principle, we can calculate the double-folding potential directly by using 
the nucleon density calculated with the AMD. The nucleon density is, however, 
inaccurate in the asymptotic region, 
since each nucleon is described by a Gaussian wave packet in the AMD. 
 Very lately we proposed a way of making a tail correction 
 to the AMD density~\cite{Minomo:2011bb}.  
 Although the calculation based on the resonating group method 
 is quite time-consuming, it was applied 
 to $^{31}$Ne~\cite{Minomo:2011bb}, since 
 the correction is most significant for $^{31}$Ne that 
 is a most-weakly bound system among $^{20-32}$Ne. 
 The tail correction to $\sigma_{\rm R}$ is about 3~\% for $^{31}$Ne. 
 The DFM with the tail-corrected density reproduces 
 the measured $\sigma_{\rm R}$ for $^{31}$Ne, whereas the DFM 
 without the tail correction underestimates the data slightly.

In this paper, we analyze the reaction cross section for the scattering 
of $^{20-32}$Ne from a $^{12}$C target at 240~MeV/nucleon 
by using the DFM with the Melbourne $g$-matrix 
in order to determine deformations of $^{20-32}$Ne systematically. 
Here the projectile density is constructed either (I) 
by the AMP-AMD calculation with the Gogny D1S interaction
or (II) by the Woods-Saxon model with the deformation obtained by the
AMP-AMD calculation. Model I has no adjustable 
parameter, but the density is inaccurate in the asymptotic region. 
Model II provides the nucleon density with the proper asymptotic form, 
but the model includes potential parameters. 
As the potential parameter set, we use the parameter set recently proposed 
by R.~Wyss~\cite{WyssPriv}.
This set is intended to reproduce the spectroscopic properties of 
high-spin states from light to heavy deformed nuclei,
e.g., the quadrupole moments and the moments of inertia,
and at the same time the RMS radii crucial for the present analysis.

Models I and II yield almost the same $\sigma_{\rm R}$ 
for $^{24-29}$Ne that have large one-neutron separation energies. 
Furthermore, this agreement is seen for $^{31}$Ne, when 
the tail correction is made in Model I. 
This indicates that Model II is a handy way of simulating 
results of Model I with the tail correction. 
Model II is quite practical compared with Model I with the tail correction 
that requires time-consuming calculations. 
Model I with the tail correction and Model II reproduce the measured 
$\sigma_{\rm R}$ of $^{20,28-32}$Ne. The deformation of
$^{28-32}$Ne is then definitely determined through this analysis. 
This analysis also yields a reasonable prediction for the deformation of 
$^{21-27}$Ne. We also confirm that $^{31}$Ne is a halo nucleus 
with large deformation. 
Furthermore, we analyze the AMP effect and the pairing effect 
on $\sigma_{\rm R}$ by using Model II.

The theoretical framework is presented in Sec.~\ref{Theoretical framework}.
We explain the DFM, the AMP-AMD and the Woods-Saxon mean-field model.
We present a handy way of making a center-of-mass (CM) correction
to the density of the mean-field model. 
We also show in Sec.~\ref{Theoretical framework} 
that the dynamical deformation effect and 
the reorientation effect neglected in the present DFM are small. 
This indicates that $\sigma_{\rm I} \approx \sigma_{\rm R}$. 
Numerical results are shown in Sec.~\ref{Results}.
Section \ref{Summary} is devoted to summary.

\section{Theoretical framework}
\label{Theoretical framework}

\subsection{Double folding model}

We start with the many-body Schr\"odinger equation
with the realistic NN interaction $v_{ij}$
for the scattering of projectile (P) on target (T):
\bea
(T_R +h_{\rm P}+h_{\rm T}+ \sum_{i \in {\rm P}, j \in {\rm T}} v_{ij}-E){\Psi}^{(+)}=0  \;,
\label{schrodinger-bare}
\eea
where
$E$ is the energy of the total system,
$T_R$ is the kinetic energy of relative motion
between P and T, and
$h_{\rm P}$ ($h_{\rm T}$) is the internal Hamiltonian of P (T). 
The multiple-scattering theory~\cite{Watson, KMT} 
for nucleon-nucleus scattering 
was extended to nucleus-nucleus scattering~\cite{Yahiro-Glauber}. 
According to the theory, Eq. \eqref{schrodinger-bare} is 
approximated into 
\bea
(T_R +h_{\rm P}+h_{\rm T}+ \sum_{i \in {\rm P}, j \in {\rm T}}
\tau_{ij}-E){\hat \Psi}^{(+)}=0  \;,
\label{schrodinger-effective}
\eea
where $\tau_{ij}$ is the effective NN interaction in the nuclear medium. 
The Brueckner $g$ matrix has often been used as such $\tau_{ij}$ 
in many applications;  see for example 
Refs.~\cite{M3Y,JLM,Brieva-Rook,Satchler-1979,Satchler,CEG,
Rikus-von Geramb,Amos,CEG07}.
The $g$ matrix interaction includes the nuclear-medium effect, 
but not the effect of collective excitations induced by 
the surface vibration and the rotation of finite nucleus, 
since the interaction is evaluated in the nuclear matter. 
The effect of collective excitations is small as shown 
in Sec.~\ref{Reorientation effect}.

In the scattering analyzed here, furthermore, 
the projectile-breakup effect is small, 
since the target is light and $E$ is large. 
This is explicitly shown in Sec.~\ref{AMD analysis for Ne isotopes}. 
In this situation the DFM becomes reliable. 
In the model, the potential $U$ between P and T  consists of the direct and exchange parts, $U^{\rm D}$ and $U^{\rm EX}$, defined by
~\cite{DFM-standard-form,DFM-standard-form-2}
\bea
\label{eq:UD}
U^{\rm DR}(\vR) \hspace*{-0.15cm} &=& \hspace*{-0.15cm} 
\sum_{\mu,\nu}\int \rho^{\mu}_{\rm P}(\vrr_{\rm P}) 
            \rho^{\nu}_{\rm T}(\vrr_{\rm T})
            g^{\rm DR}_{\mu\nu}(s;\rho_{\mu\nu}) d \vrr_{\rm P} d \vrr_{\rm T}, \\
\label{eq:UEX}
U^{\rm EX}(\vR) \hspace*{-0.15cm} &=& \hspace*{-0.15cm}\sum_{\mu,\nu} 
\int \rho^{\mu}_{\rm P}(\vrr_{\rm P},\vrr_{\rm P}+\vs)
\rho^{\nu}_{\rm T}(\vrr_{\rm T},\vrr_{\rm T}-\vs) \nonumber \\
            &&~~\hspace*{-0.5cm}\times g^{\rm EX}_{\mu\nu}(s;\rho_{\mu\nu}) \exp{[i\vK(\vR) \cdot \vs/M]}
            d \vrr_{\rm P} d \vrr_{\rm T},~~~~
            \label{U-EX}
\eea
where $\vs=\vrr_{\rm P}-\vrr_{\rm T}+\vR$ for the coordinate $\vR$
of P from T. Indices $\mu$ and $\nu$ stand for the $z$-component
of isospin ; $\mu = 1/2 (\nu = 1/2) $ means neutron and $\mu = -1/2 (\nu = -1/2)$ does proton.
The original form of $U^{\rm EX}$ is a non-local function of $\vR$,
but  it has been localized in Eq.~\eqref{U-EX}
with the local semi-classical approximation~\cite{Brieva-Rook} in which
P is assumed to propagate as a plane wave with
the local momentum $\hbar \vK(\vR)$ within a short range of the
 NN interaction, where $M=A A_{\rm T}/(A +A_{\rm T})$
for the mass number $A$ ($A_{\rm T}$) of P (T).
The validity of this localization is shown in Ref.~\cite{Minomo:2009ds}.

The direct and exchange parts, $g^{\rm DR}_{\mu\nu}$ and 
$g^{\rm EX}_{\mu\nu}$, of the effective NN ($g$-matrix) interaction are 
assumed to depend on the local density
\bea
 \rho_{\mu\nu}=\rho^{\mu}_{\rm P}(\vrr_{\rm P}+\vs/2)
 +\rho^{\nu}_{\rm T}(\vrr_{\rm T}-\vs/2)
\label{local-density approximation}
\eea
at the midpoint of the interacting nucleon pair.
The direct and exchange parts, $g^{\rm DR}_{\mu\nu}$ and $g^{\rm EX}_{\mu\nu}$,
are described by
\bea
g_{\mu,\nu,T_{z}=\pm1}^{\rm DR}(s;\rho_{\mu\nu}) \hspace*{-0.15cm} &=&
 \hspace*{-0.15cm}
\frac{1}{4} \sum_S \hat{S}^2 g_{\mu\nu}^{S1} (s;\rho_{\mu\nu})
\delta^{\mu+\nu}_{T_z} , \\
g_{\mu,\nu,T_{z}=0}^{\rm DR}(s;\rho_{\mu\nu}) \hspace*{-0.15cm} &=&
 \hspace*{-0.15cm}
\frac{1}{8} \sum_{S,T} \hat{S}^2 g_{\mu\nu}^{ST}(s;\rho_{\mu\nu}) 
\delta^{\mu+\nu}_{T_z} , \\
g_{\mu,\nu,T_{z}=\pm1}^{\rm EX}(s;\rho_{\mu\nu})\hspace*{-0.15cm}  &=&
 \hspace*{-0.15cm}
\frac{1}{4} \sum_S (-1)^{S+1} \hat{S}^2 g_{\mu\nu}^{S1} (s;\rho_{\mu\nu})
\delta^{\mu+\nu}_{T_z} , \\
g_{\mu,\nu,T_{z}=0}^{\rm EX}(s;\rho_{\mu\nu})\hspace*{-0.15cm} &=&
 \hspace*{-0.15cm}
\frac{1}{8} \sum_{S,T} (-1)^{S+T} \hat{S}^2 g_{\mu\nu}^{ST}(s;\rho_{\mu\nu})
\delta^{\mu+\nu}_{T_z},~~~
\eea
with $\hat{S} = {\sqrt {2S+1}}$, 
in terms of the spin-isospin components $g_{\mu\nu}^{ST}$ of the $g$ matrix
interaction.
As for the $g$ matrix interaction, we take
a sophisticated version of the Melbourne interaction
\cite{Amos} that is constructed from the Bonn-B 
NN potential \cite{BonnB}. 
In actual calculations, the relativistic kinematics is taken for $T_R$
and $E$.

\subsection{AMD framework and inputs for the reaction calculations}
The framework and calculational procedure of the AMD 
in this study are common to those of
Ref. \cite{Kimura1}, and the reader is directed to it for more detail.
The Hamiltonian of the AMD is given as
\begin{align}
 H&= T_{\rm tot} + \sum_{i<j} {\bar v}_{ij} - T_{\rm cm}.
\end{align}
The Gogny D1S \cite{GognyD1S}
is used as an effective nucleon-nucleon interaction
${\bar v}_{ij}$; here note that the Coulomb part of ${\bar v}_{ij}$ 
is approximated 
by a sum of twelve Gaussians. $T_{\rm tot}$ and $T_{\rm cm}$
represent the kinetic energies of nucleons and center-of-mass motion,
respectively.

The variational wave function is a parity-projected wave function and
the intrinsic wave function is a Slater determinant of nucleon wave packets,
\begin{align}
 \Phi^\pi&=P^\pi{\cal A}\left\{\varphi_1,\varphi_2,...,\varphi_A
 \right\},
 \label{eq:amdint}
\end{align}
where $P^\pi$ is the parity projector. The nucleon wave packet $\varphi_i$
is a direct product of spatial $\phi_i$, spin $\chi_i$
and isospin $\xi_i$ parts,
\begin{align}
 \varphi_i&=\phi_i(\bm r)\chi_i\xi_i,\\
 \phi_i(\bm r) &= \prod_{\sigma=x,y,z}\left(\frac{2\nu_\sigma}{\pi}\right)^{1/4}
 \exp\left\{-\nu_\sigma\left(r_{\sigma} - \frac{Z_{i\sigma}}{\sqrt{\nu_\sigma}}\right)^2\right\}, \\
 \chi_i &= \alpha_{i,\uparrow} \chi_{\uparrow} + \alpha_{i,\downarrow}
 \chi_{\downarrow} ,
 \quad \xi_i = \text{p or n},
\end{align}
where the centroids of Gaussian wave packets $\bm Z_i$, the width of Gaussian
$\nu_\sigma$ and the spin direction $\alpha_{i,\uparrow}$ and
$\alpha_{i,\downarrow}$ are the parameters determined
variationally as explained below. Note that the center-of-mass wave function is analytically
separable from the variational wave function Eq. (\ref{eq:amdint}):
\begin{align}
 \Phi^\pi&=\Phi_{\rm cm}\Phi_{\rm int},\\
 \Phi_{\rm cm} &= \prod_{\sigma=x,y,z}
 \left(\frac{2A\nu_\sigma}{\pi}\right)^{1/4} \hspace*{-0.5cm}
 \exp\left\{-A\nu_\sigma\left(X_{\sigma}-\frac{Z_{\rm cm,\sigma}}{\sqrt{A\nu_\sigma}}\right)^2\right\}, \\
 \bm Z_{\rm cm} &= \frac{1}{\sqrt{A}}\sum_{i=1}^A\bm Z_i,
 \label{ZCM}
\end{align}
where $\bm X$ represents the center-of-mass coordinate.
Usually, $\bm Z_{\rm cm}$ defined by Eq.~\eqref{ZCM} can be
set to zero without loss of generality.
This is common to the angular momentum projection and GCM calculation,
and all quantities used as inputs of reaction calculation are free from the
spurious center-of-mass motion.

The parameters in Eq. (\ref{eq:amdint}) are determined using the frictional cooling method
to minimize the total energy under the constraint on the matter quadrupole deformation
parameter $\bar{\beta}$.  Here the quadrupole deformation parameters are defined as,
\begin{align}
 \frac{\left< x^2 \right>^{1/2}}
 {\left[ \left< x^2 \right> \left< y^2 \right> \left< z^2 \right>
 \right]^{1/6}} &= \exp \left[ {\sqrt {\frac{5}{4\pi}}} \bar{\beta} \cos
 \left( \bar{\gamma} + \frac{2\pi}{3} \right) \right],
 \label{eq:AMDbgx2}\\
 \frac{\left< y^2 \right>^{1/2}}
 {\left[ \left< x^2 \right> \left< y^2 \right> \left< z^2 \right>
 \right]^{1/6}} &= \exp \left[ {\sqrt {\frac{5}{4\pi}}} \bar{\beta} \cos
 \left( \bar{\gamma} - \frac{2\pi}{3} \right) \right],
 \label{eq:AMDbgy2}\\
 \frac{\left< z^2 \right>^{1/2}}
 {\left[ \left< x^2 \right> \left< y^2 \right> \left< z^2 \right>
 \right]^{1/6}} &= \exp \left[ {\sqrt {\frac{5}{4\pi}}} \bar{\beta} \cos
 \bar{\gamma}  \right].
 \label{eq:AMDbgz2}
\end{align}
Here, $\left< x^2 \right>$, $\left< y^2 \right>$ and
$\left< z^2 \right>$ are calculated from $\Phi_{\rm int}$ in the  intrinsic frame
that is so chosen to satisfy the relation
$\left< x^2 \right> \le \left< y^2 \right> \le\left< z^2 \right>$.
The constraint is imposed on the value of  $\bar{\beta}$ from 0 to 1.0 with the
interval of 0.025. Since we do not make any assumption on the spatial symmetry of the wave
function and do not impose any constraint on $\bar{\gamma}$, it has an 
optimal value for each given value of $\bar{\beta}$.

After the variation, we perform the angular momentum projection for each value of $\bar{\beta}$,
\begin{align}
 \Phi^{J\pi}_{MK}(\bar{\beta})&=P^{J}_{MK}\Phi_{\rm int}^\pi(\bar{\beta}),\label{eq:amdproj}\\
 P^{J\pi}_{MK} &= \frac{2J+1}{8\pi^2}\int d\Omega D^{J*}_{MK}(\Omega)R(\Omega),\label{eq:jproj}
\end{align}
where $D^J_{MK}(\Omega)$ and $R(\Omega)$ are Wigner's $D$ function and rotation operator,
respectively. The integrals over three Euler angles $\Omega$ in Eq. (\ref{eq:jproj}) are
performed numerically.

The AMD calculation is completed by performing the GCM. The wave functions that have the same
parity and angular momentum $(J, M)$ are superposed as
\begin{align}
 \Phi_n^{J\pi}&= \sum_{K=-J}^{J} \sum_{\bar{\beta}} c_{nK} (\bar{\beta})
 \Phi_{MK}^{J\pi}(\bar{\beta}) .
\end{align}
In other words, $K$ and $\bar{\beta}$ are the generator coordinates in this calculation. The
coefficients $c_{nK}(\bar{\beta})$ are determined by solving the Hill-Wheeler equation.
\begin{align}
 \sum_{\bar{\beta}^\prime K^\prime}
 H_{K K^\prime}(\bar{\beta},\bar{\beta}^{\prime})
 c_{n K^\prime} (\bar{\beta}^{\prime}) =
 E_n \sum_{\bar{\beta}^\prime K^\prime}
 N_{K K^\prime} (\bar{\beta},\bar{\beta}^{\prime})
 c_{n K^\prime} (\bar{\beta}^{\prime}),\\
\left\{
 \begin{array}{c}
  N_{K K^\prime} (\bar{\beta},\bar{\beta}^{\prime})\\
  H_{K K^\prime} (\bar{\beta},\bar{\beta}^{\prime})
 \end{array}
 \right\} =
 \langle \Phi_{MK}^{J\pi}(\bar{\beta})|
 \left\{
 \begin{array}{c}
  1\\
  H
 \end{array}
 \right\}
 | \Phi^{J\pi}_{MK^\prime}(\bar{\beta}^{\prime})\rangle.
\end{align}
The ground state wave function $\Phi^{J\pi}_{\rm g.s.}$ obtained by this procedure is used in the discussion of Sec. \ref{Results}.

For the reaction calculation, two types of applications are performed.
One is to use the deformation parameters $\bar{\beta}$ and $\bar{\gamma}$
as inputs of deformed Woods-Saxon potential.
We assign the deformation of the AMD wave function by picking 
up a GCM basis wave function that has the
maximum overlap with the ground state wave function,
$| \langle \Phi^{JM\pi}_{\rm g.s.}|\Phi^{J\pi}_{MK}(\bar{\beta})\rangle|^2$.
And then we define the deformation of AMD wave function as equal to
$\bar{\beta}$ and corresponding $\bar{\gamma}$.

The other is the direct use of the nucleon density  calculated from
the ground state wave function as an input of the double-folding potential,
\begin{align}
 \rho_{JMJM'}(\bm r)&=\langle \Phi^{JM\pi}_{\rm g.s.}|\sum_{i}
 \delta(\bm r_i - \bm X - \bm r)|\Phi^{JM'\pi}_{\rm g.s.}\rangle,
 \label{eq:amddens1}\\
 &=\sum_{\lambda=0}^{2J} \rho_{JJ}^{(\lambda)}(r)
 (JM' \lambda \mu |JM)
 Y^*_{\lambda \mu}(\hat r),
 \label{eq:amddens}
\end{align}
where the summation of $\lambda$
in Eq. (\ref{eq:amddens}) runs for even numbers.
When only $\rho_{JJ}^{(\lambda=0)}$ is taken in the double-folding potential,
the resultant folding potential becomes spherical. This approximation is
often used as a standard manner in the DFM.
The validity of this approximation is shown 
in Sec.~\ref{Reorientation effect}.

\subsection{Woods-Saxon mean-field model}

We also perform DFM calculations with resultant density calculated
by the simple mean-field model based on the Woods-Saxon potential.
Sophisticated AMD calculation is a powerful tool but
it costs much time to obtain reliable information, so, we take a deformed
Woods-Saxon model for alternative approach and for further
investigation. As shown later, in many cases of Ne isotope induced
reaction cross section, the deformed Woods-Saxon model gives results
consistent with the AMD calculation if one uses the consistent deformation.

The deformed Woods-Saxon potential in Schr\"odinger equation
is composed of the central and spin-orbit part,
which have the following forms:
\begin{align}
\label{DWS-0}
   V_{\rm c}( {\bfi r})
 = \frac{V_0}{1+\exp \left[ {\rm dist}_{\Sigma}
   ({\bfi r})/a \right]},
\end{align}
\begin{align}
\label{DWS-1}
   V_{\rm so}( {\bfi r})
 &= \lambda_{\rm so}\left(\frac{\hbar}{2m_{\rm red}c}\right)^2
  \bm{\nabla}V_{\rm c}(\bm{r})\cdot
  \left(\bm{\sigma}\times \frac{1}{i}\bm{\nabla}\right), 
\end{align}
where 
$m_{\rm red}=m(A-1)/{A}$ and 
${\rm dist}_{\Sigma} ({\bfi r})$ represents a distance between 
a given point ${\bfi r}$ and 
the deformed surface $\Sigma$ specified by the radius,
\begin{align}
 R(\theta,\phi;\bm{\alpha}) = R_{0}c_{v}(\bm{\alpha})\Bigr[
 1+\sum_{\lambda\mu}\alpha^*_{\lambda\mu}Y_{\lambda\mu}(\theta,\phi)\Bigl],
\label{eq:surf}
\end{align}
with the deformation parameters
$\bm{\alpha} \equiv \left\{ \alpha_{\lambda\mu}\right\} $.
The constant $c_v (\bm{\alpha})$ is introduced to guarantee
the volume conservation of nucleus.
A set of deformation parameters used in the present work is
$(\beta_2,\gamma,\beta_4)$~\cite{NS81},
which are related to $(\alpha_{2\mu},\alpha_{4\mu})$ by
\begin{equation}
\left\{ \begin{array}{l}
 \alpha_{20}=\beta_2\cos\gamma,\\
 \alpha_{22}=\alpha_{2-2}=-\frac{1}{\sqrt{2}}\,\beta_2\sin\gamma,\\
 \alpha_{40}=\frac{1}{6}\,\beta_4
  (5\cos^2\gamma+1),\\
 \alpha_{42}=\alpha_{4-2}=-\sqrt{\frac{5}{6}}\,\beta_4
  \cos\gamma\sin\gamma,\\
 \alpha_{44}=\alpha_{4-4}=\sqrt{\frac{35}{72}}\,\beta_4
  \sin^2\gamma, 
\end{array} \right.
\label{eq:WSbeta}
\end{equation}
where note that the other $\alpha_{\lambda\mu}$ are zero.

As for the parameter set of the Woods-Saxon potential, i.e.,
the potential depth $V_0$, the nuclear radius $R_0$, and
the diffuseness parameter $a$ of the central potential,
as well as those for the spin-orbit potential,
we take the one provided recently by R. Wyss~\cite{WyssPriv}
(see Ref.~\cite{SS09} for the actual values of parameters).
For proton, the Coulomb potential created by charge $(Z-1)e$ that has a 
uniform distribution inside the surface $\Sigma$ is added to
Eq.~(\ref{DWS-0}); more detailed description are explained in e.g.
Refs.~\cite{CDN87}.

The deformation parameters in the Woods-Saxon potential
are determined by the standard Strutinsky
(microscopic-macroscopic) method~\cite{Str68,FunnyHills},
where the pairing correlation is included within the BCS approximation.
The monopole pairing interaction is used and its strength
is determined according to the smoothed pairing gap method.
As for the macroscopic part energy, the liquid-drop model
of Ref.~\cite{MS67} is employed.
The Ne isotopes in the IOI region studied in the present work
are sitting near the drip line.  In such a case the standard Strutinsky
method has problems related to the continuum single-particle states.
Recently the problems are solved by using the so-called
Kruppa prescription~\cite{TST10}.  We have employed this improved
method for the calculation of both the shell correction and
the pairing correlation (Kruppa-BCS method).

As it is discussed in the following sections,
we utilize different models for the analysis of reaction cross sections.
In order to compare the deformation in different models,
it is necessary to transform the deformation parameters defined
within each model, e.g., between $(\bar{\beta},\bar{\gamma})$
in the AMD model and $(\beta_2,\gamma,\beta_4)$ in the Woods-Saxon model.
This has been done in the following way.
The deformed surface in Eq.~(\ref{eq:surf}) defines the deformation parameter
in the Woods-Saxon potential.
We define the uniform density with the sharp cut surface $\Sigma$,
\begin{equation}
 \rho_{\rm uni}(\bm{r})\equiv\rho_0\theta(R(\theta,\phi;\bm{\alpha})-r),
\label{eq:unirho}
\end{equation}
where $\rho_0$ is the average density and  $\theta(x)$ is a step function,
and calculate the expectation value with it, e.g.,
\begin{equation}
 \langle x^2 \rangle_{\rm uni} =\int x^2 \rho_{\rm uni}(\bm{r})d\bm{r}.
\label{eq:x2uni}
\end{equation}
For a given AMD deformation parameters $(\bar{\beta},\bar{\gamma})$,
the ratio of the AMD expectation values
$\langle x^2 \rangle:\langle y^2 \rangle:\langle z^2 \rangle$ is
fixed according to Eqs.~(\ref{eq:AMDbgx2})$-$(\ref{eq:AMDbgz2}).
Then the corresponding Woods-Saxon parameters $(\beta_2,\gamma,\beta_4)$
are defined to give the same shape, i.e, by the condition
$\langle x^2 \rangle:\langle y^2 \rangle:\langle z^2 \rangle
=\langle x^2 \rangle_{\rm uni}:\langle y^2 \rangle_{\rm uni}:
\langle z^2 \rangle_{\rm uni}$.  This condition gives only
two independent equations so that the $(\beta_2,\gamma)$ is determined
under some fixed value of $\beta_4$.
We set $\beta_4=0$ for simplicity in order to define
the $(\beta_2,\gamma)$ values corresponding to the AMD calculation.
It is then found that the pairs $(\bar{\beta},\bar{\gamma})$
and $(\beta_2,\gamma)$ take similar values; 
see Table~\ref{tab:DWS-parameter} in Sec.~\ref{Results}.
We have checked that the results of the final reaction cross sections
change very little (order of few mb),
if we use the non-zero $\beta_4$ values within the range
$-0.1 < \beta_4 < 0.1$.

In the actual calculation, the Woods-Saxon potential is diagonalized
with the anisotropic harmonic oscillator basis,
where the three frequencies, $\omega_i$ ($i=x,y,z$),
are taken to be proportional to $1/\sqrt{\langle x_i^2 \rangle_{\rm uni}}$,
which is close to the optimal choice.
As for the basis size, we have used the oscillator shells
$N_{\rm osc}=n_x+n_y+n_z = 18$ in most cases.
However, the density distribution of the nucleus near the drip line
extends considerably, and then
we have checked the convergence of the results carefully by taking
larger number of shells $N_{\rm osc}\ge 20$ in such a case.

As in the Hartree-Fock (HF) or Hartree-Fock-Bogoliubov (HFB) approach,
the occurrence of deformation in the Woods-Saxon model is
a symmetry-breaking phenomenon.
The many-body wave function $\Phi$
is then considered to be that in the intrinsic (body-fixed) frame~\cite{RS80},
and so is the nucleon density calculated with $\Phi$,
\begin{equation}
 \rho^{(\rm in)}(\bm{r})=
 \langle\Phi|\sum_i\delta(\bm{r}_i-\bm{r})|\Phi\rangle
 =\sum_\alpha |\varphi_\alpha(\bm{r})|^2v^2_\alpha,
\end{equation}
where $\varphi_\alpha(\bm{r})$ is the Woods-Saxon single-particle wave function
and $v_\alpha$ is the BCS occupation probability
(the free contributions should be subtracted when the Kruppa prescription
is employed, see Refs.~\cite{TST10,OST10} for details).
Therefore the deformed density
$\rho^{(\rm in)}(\bm{r})=\rho^{(\rm in)}(r,\theta,\phi)$ cannot be
directly used in the reaction calculation such as the DFM that is done
in the laboratory (space-fixed) frame.

\begin{figure}[htbp]
\begin{center}
 \includegraphics[width=0.43\textwidth,clip]{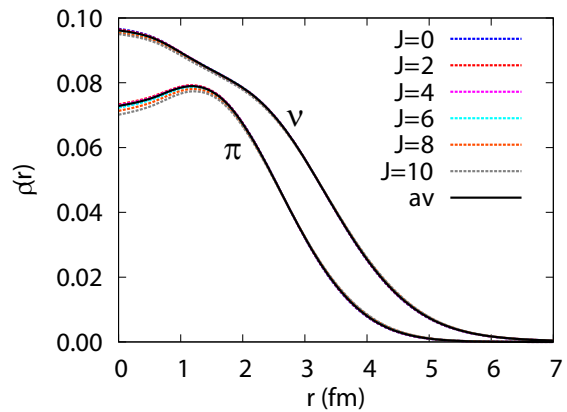}
 \caption{(Color online)
 Comparison of the projected density,
 $\rho_{JJ}^{(0)}(r)/\sqrt{4\pi}$ with $J=0,2,...,10$,
 and the angle-averaged intrinsic density,
 $\rho_{\rm av}^{\rm (in)}(r)$ in Eq.~(\ref{eq:avdens}),
 for the Woods-Saxon model in $^{30}$Ne.
 The deformation parameters are $\beta_2=0.4$ and $\gamma=\beta_4=0$,
 and the pairing gaps $\Delta_{\rm n}=\Delta_{\rm p}=1$ MeV.
 The results are very similar also for the case with no pairing correlation. 
  }   
 \label{Fig-proj-dens}
\end{center}
\end{figure}

One way to recover the spherical symmetry and transform the density
in the intrinsic frame to that in the laboratory frame is to perform
the angular momentum projection,
as already explained in Eqs.~(\ref{eq:amddens1})$-$(\ref{eq:amddens}) 
in the AMD framework.
We have performed the projection calculation 
(without the CM correction, which is discussed in the next subsection)
by using the method of Ref.~\cite{TStobeP}
for the Woods-Saxon model with the BCS pairing correlation.
It is found that the projected density $\rho_{JJ}^{(0)}(r)/\sqrt{4\pi}$,
where $\rho_{JJ}^{(\lambda=0)}(r)$ is defined in the same way
as in Eq.~(\ref{eq:amddens}) and used in the DFM,
is very similar to the following angle-averaged intrinsic density,
\begin{equation}
 \rho_{\rm av}^{(\rm in)}(r)=
 \frac{1}{4\pi}\int\rho^{\rm (in)}(r,\theta,\phi)\sin{\theta}d\theta d\phi.
\label{eq:avdens}
\end{equation}
In Fig.~\ref{Fig-proj-dens},
they are compared for $^{30}$Ne.  As it is clear, for both neutron and
proton, the projected and angle-average densities are almost identical
particularly in the tail region,
while slight differences are observed in the inner region
for large $J$ values.  This is rather general trend, and therefore,
we use $\rho_{\rm av}^{(\rm in)}(r)$ in place of
$\rho_{JJ}^{(0)}(r)/\sqrt{4\pi}$
in the Woods-Saxon model for the ground state of Ne isotopes.

In the view of static behaviour of deformed nuclei, using
spherical part of deformed Woods-Saxon density,
$\rho_{JJ}^{(0)}(r)/\sqrt{4\pi} \approx \rho_{\rm av}(r)$,
is well justified, but,
this procedure doesn't justify that dynamical aspect of deformation
is also negligible. As mentioned above, this point is argued in 
Sec.~\ref{Reorientation effect}.

\subsection{CM correction to nucleon density of
the Woods-Saxon mean-field model}
\label{Section:CM-correction}

The projectile density is constructed with either the AMD 
or the Woods-Saxon model for Ne isotopes.
In contrast to the AMD calculation,
the CM motion is not excluded
in the many-body wave function $\Phi$ in the Woods-Saxon model.
We then extract the CM motion from $\Phi$
in the standard manner~\cite{TB-1958,Suzuki-C75} and
propose a simple extraction prescription consistent with the standard manner.

The wave function $\Phi$ is approximated by a product of
the CM part $\Phi_{\rm cm}$ and the intrinsic part $\Phi_{\rm int}$:
\bea
\Phi=\Phi_{\rm cm}\Phi_{\rm int}
\eea
with
\bea
\Phi_{\rm cm}=\Big( \frac{A}{\pi b^2} \Big)^{3/4}
\exp{\Big[-\frac{A}{2b^2}X^2\Big]}
\eea
for the CM coordinate $\vX$ and the size parameter $b$.
The mean squared radii of $\Phi$ and
$\Phi_{\rm int}$ are obtained by
\bea
\langle \vrr^2 \rangle &\equiv&
\langle \Phi |\sum_{i} \vrr_{i}^2 |\Phi \rangle ,
\\
\langle \vrr^2 \rangle_{\rm int} &\equiv&
\langle \Phi_{\rm int} |\sum_{i} (\vrr_{i}-\vX)^2 |\Phi_{\rm int} \rangle
\eea
for a single-particle coordinate $\vrr_{i}$, and hence 
these are related to $b$ as
\bea
\langle \vrr^2 \rangle =
\langle \vrr^2 \rangle_{\rm int} + \frac{3}{2}\frac{b^2}{A}.
\label{r2-relation}
\eea
The CM correction to $\langle \vrr^2 \rangle$ is small (order $1/A$),
so it can be estimated with $\Phi$ instead of $\Phi_{\rm int}$:
\bea
\langle \vrr^2 \rangle_{\rm int} \approx
\langle \Phi |\sum_{i} (\vrr_{i}-\vX)^2 |\Phi \rangle.
\label{r2-approx}
\eea
The correction is a combination of the one-body and two-body corrections.
Inserting Eq.~\eqref{r2-approx} into Eq.~\eqref{r2-relation},
we can determine the size parameter $b$ and then $\Phi_{\rm int}$ from $\Phi$.

The proton and neutron density without and with the CM correction are
obtained by
\bea
\rho(\vrr) &=& \langle \Phi |\sum_{i} \delta(\vrr_{i}-\vrr) P_{i}
|\Phi \rangle ,
\\
\rho_{\rm int}(\vrr) &=&
\langle \Phi_{\rm int} |\sum_{i} \delta(\vrr_{i}-\vX-\vrr)
P_{i}|\Phi_{\rm int} \rangle,
\eea
where $P_{i}$ is a projector for proton or neutron. These densities satisfy
\bea
\rho(\vrr)=\int d\vrr' |\Phi_{\rm cm}(\vrr-\vrr')|^2 \rho_{\rm int}(\vrr').
\label{CM-folding}
\eea
Thus, the density $\rho_{\rm int}$ with the CM correction is obtained by
unfolding $\rho$ with $|\Phi_{\rm cm}|^2$.

Instead of the complicated unfolding procedure~\cite{TB-1958}, we can take
the following simple prescription.
As shown in Eq. \eqref{r2-relation}, the difference between
$\langle \vrr^2 \rangle$ and $\langle \vrr^2 \rangle_{\rm int}$ is small,
because it is of order $1/A$. This indicates that $\vrr$ dependence of
$\rho_{\rm int}(\vrr)$ is similar to that of $\rho(\vrr)$.
We can then approximate $\rho_{\rm int}(\vrr)$ by
\bea
\rho_{\rm int}(\vrr)=\frac{1}{\a^3}\rho(\vrr/\a)
\eea
with a scaling factor
\bea
\a=\sqrt{ \frac{\langle \vrr^2 \rangle_{\rm int}}{\langle \vrr^2 \rangle}}
=\sqrt{1-\frac{3}{2A}\frac{b^2}{\langle \vrr^2 \rangle}} ,
\eea
where $\a$ has been determined to reproduce $\langle \vrr^2 \rangle_{\rm int}$
of Eq. \eqref{r2-approx}.
The error of this simple prescription to the unfolding procedure is
only 0.1\% in $\sigma_{\rm R}$ for Ne isotopes, so we
use the simple prescription for the density calculated with
the mean-field model.
The RMS radii,
$\sqrt{\langle \vrr^2 \rangle}$ and $\sqrt{\langle \vrr^2 \rangle_{\rm int}}$,
without and with the CM correction
are estimated with the spherical HF model, and
the parameter $b$ is evaluated with Eq. \eqref{r2-relation}
from the RMS radii.
For each of Ne isotopes, we use a common $b$ among
the HF calculation and the spherical and deformed WS calculations, since
the difference of $\sqrt{\langle \vrr^2 \rangle}$ among
these mean-field models are at most 6\% and the 6\% error to the 1.5\% CM correction is negligible.

\subsection{Dynamical deformation and reorientation effects}
\label{Reorientation effect}

When the projectile is deformed in the intrinsic frame, 
the deformation enlarges the radius of the projectile density 
in the space-fixed frame 
and eventually enhances the reaction cross section. 
This static deformation effect has already been included 
in the DFM by making the AMP. 
Another effect to be considered is the dynamical deformation effect 
that is an effect of the rotational motion of the deformed projectile 
during the scattering. This effect on the reaction cross section 
is found to be small for intermediate-energy nucleus-nucleus 
scattering~\cite{Minomo-DWS}. This was confirmed with the adiabatic 
approximation to the rotational motion of projectile and 
the eikonal approximation to the relative motion between the projectile and 
the target. In this subsection, the effect is investigated 
with no approximation.

In order to test the dynamical deformation effect, we consider 
the scattering of $^{30}$Ne from $^{12}$C at 240~MeV/nucleon and 
do a coupled-channel calculation between the $0^{+}$ ground state 
and the  first $2^{+}$ state of $^{30}$Ne. 
The projectile density is calculated by the DWS model 
with the deformation evaluated by the AMD. The coupling potentials 
in the coupled-channel calculation 
are obtained by the so-called single-folding model; namely, 
the nucleon-$^{12}$C potential is first evaluated by folding 
the Melbourne-$g$ matrix interaction with the target density of $^{12}$C and 
the coupling potentials are then obtained by folding 
the nucleon-$^{12}$C potential with the projectile transition densities.

In the single-channel calculation with no dynamical deformation effect, 
the resultant reaction cross section is 1469~mb. 
This result overestimates the corresponding result of the DFM by about
$10~\%$, which is enough to accuracy of the present test. 
In the coupled-channel calculation with the dynamical deformation effect 
from the first $2^{+}$ state, the resulting reaction cross section is 1468~mb. 
Thus the dynamical rotation effect on the reaction cross section is 
estimated as less than $0.1~\%$.  
The reason why the effect is small 
for intermediate-energy nucleus-nucleus scattering is shown in 
Ref.~\cite{Minomo-DWS}. 
The integrated inelastic cross section to the first $2^{+}$ state is 2.9~mb. 
This is 0.2~\% of $\sigma_{\rm R}$, indicating that 
$\sigma_{\rm I} \approx \sigma_{\rm R}$. 

The folding potential $U$ is not spherical in general, when 
the spin of the projectile is not zero. This effect called 
the reorientation effect is also tested by the coupled-channel
calculation for the scattering of $^{31}$Ne$(3/2^{-})$ from $^{12}$C at 
240~MeV/nucleon, where the single-folding model is used. 
The resultant reaction cross section is 1512mb, whereas 
the corresponding cross section is 1515mb 
when the non-spherical part of $U$ is switched off.
The reorientation effect is 0.2~\% and hence negligible
for intermediate-energy nucleus-nucleus scattering.

\section{Results}
\label{Results}

\subsection{Reaction cross sections for stable nuclei}

We first test the accuracy of the DFM
with the Melbourne-$g$ matrix NN interaction
for $^{12}$C scattering at incident energies ($E_{\rm in}$)
around $240A$~MeV from stable targets,
$^{12}$C, $^{20}$Ne, $^{23}$Na and $^{27}$Al.
Experimental data on $\sigma_{\rm R}$ are available for
a $^{12}$C target at $E_{\rm in} = 250.8A$~MeV
and a $^{27}$Al target at $E_{\rm in} = 250.7A$~MeV~\cite{expC12C12}.
For $^{20}$Ne and $^{23}$Na targets,
$\sigma_{\rm I}$ at $E_{\rm in}=240A$~MeV were recently deduced
from measured $\sigma_{\rm I}$
at around 1~GeV/nucleon~\cite{Ne20-sigmaI,Na23-sigmaI}
with the Glauber model~\cite{Takechi}.

For these stable nuclei, we take the phenomenological
proton-density~\cite{C12-density} deduced
from the electron scattering by
unfolding the finite-size effect of the proton charge in the
standard manner~\cite{Singhal}, and
the neutron density is assumed to have the same geometry as the
corresponding proton one, since the proton RMS radius deviates from
the neutron one only by less than 1\% in the Hartree-Fock (HF) calculation.

DFM calculations are done with three types of effective NN interactions:
the Love-Franey $t$-matrix interaction ($t_{\rm LF}$)~\cite{Love-Franey},
the Melbourne-$g$ matrix interaction ($g_{\rm MP}$)~\cite{Amos}
evaluated from the Paris realistic NN interaction~\cite{Paris} and
the Melbourne interaction ($g_{\rm MB}$)~\cite{Amos}
constructed from the Bonn-B realistic NN interaction~\cite{BonnB}.

Table~\ref{table1} shows experimental and theoretical
reaction cross sections for a $^{12}$C target
at $E_{\rm in} = 250.8A$~MeV and a $^{27}$Al target
at $E_{\rm in} = 250.7A$~MeV.
The effective interaction $t_{\rm LF}$ has no nuclear medium effect.
In this case, the theoretical reaction cross sections
overestimate the mean values of data~\cite{expC12C12} by 17\% for $^{12}$C
and by 15\% for $^{27}$Al. In the cases of $g_{\rm MP}$ and $g_{\rm MB}$
with the medium effect, meanwhile, the overestimation is only a few percent
for both $^{12}$C and $^{27}$Al.
The medium effect is thus significant, and an amount
of the effect is almost independent of the bare NN interaction taken.
As for $^{27}$Al, the reaction cross section calculated with
$g_{\rm MB}$ agrees with the mean value of data,
when the theoretical $\sigma_{\rm R}$ is multiplied by the factor $F=0.978$.


\begin{table}
\caption{
 Reaction cross sections for $^{12}$C+$^{12}$C scattering
at $E_{\rm in}=250.8A$~MeV and $^{12}$C+$^{27}$Al scattering
at $E_{\rm in}=250.7A$~MeV. Results of three types
of effective nucleon-nucleon interactions are compared with
the corresponding data~\cite{expC12C12}.
The cross sections are presented in units of mb.
}
\label{table1}
\begin{center}
\begin{tabular}{cccccc} \hline
   & Target
   & Exp.~\cite{expC12C12}
   & $t_{\rm LF}$~\cite{Love-Franey}
   & $g_{\rm MP}$~\cite{Amos}
   & $g_{\rm MB}$~\cite{Amos}
   \\ \hline 
   & $^{12}$C
   &  782.0 $\pm$ 10.0
   &  917.7
   &  793.1
   &  795.9
   \\ \hline 
   & $^{27}$Al
   &  1159.0 $\pm$ 14.0
   &  1337.5
   &  1164.9
   &  1185.2
   \\ \hline 
\end{tabular}
\end{center}
\end{table}

\begin{figure}[htbp]
\begin{center}
 \includegraphics[width=0.43\textwidth,clip]{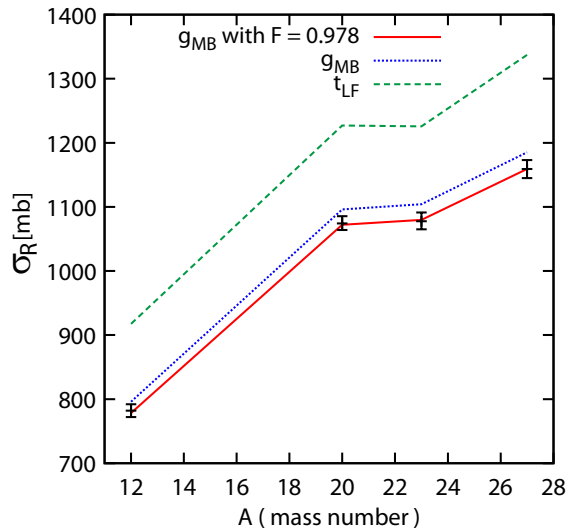}
 \caption{(Color online) Reaction cross sections for scattering of
 $^{12}$C on stable nuclei from $A=12$ to 27. The data at $250.8A$~MeV
 for $^{12}$C and $^{27}$Al are taken from Ref.~\cite{expC12C12}.
 The data at $240A$~MeV for $^{20}$Ne and $^{23}$Na are
 deduced from measured $\sigma_{\rm I}$
 at around 1~GeV/nucleon~\cite{Ne20-sigmaI,Na23-sigmaI}
 with the Glauber model~\cite{Takechi}.
 The solid (dotted) line stands for results of the DFM
 with $g_{\rm MB}$ after (before) the normalization of $F=0.978$,
 whereas the dashed line corresponds to results of $t_{\rm LF}$.
   }
 \label{Fig-reaction-Xsec-stable}
\end{center}
\end{figure}

In Fig.~\ref{Fig-reaction-Xsec-stable}, $\sigma_{\rm R}$ or $\sigma_{\rm I}$
is plotted for $^{12}$C, $^{20}$Ne, $^{23}$Na and $^{27}$Al targets.
The dotted and solid lines represent results of the DFM with $g_{\rm MB}$
before and after the normalization of $F=0.978$, respectively.
Before the normalization procedure, the dotted line slightly
overestimates the mean values of data for $A=20-27$.
After the normalization procedure, the solid line agrees
with the mean values of data for all the targets
The normalization procedure is thus reliable.
The dashed line corresponds to results of $t_{\rm LF}$ with no normalization.
The medium effect reduces the theoretical reaction cross sections
by about 15\% for all the targets.

As for the scattering of Ne isotopes on $^{12}$C at 240~MeV/nucleon, 
we perform the DFM calculation with $g_{\rm MB}$ 
and the normalization factor $F$. 
The DFM calculation with $g_{\rm MB}$ has harmless numerical ambiguity 
due to the parameterization of $g_{\rm MB}$; 
the imaginary part of the folding potential has a small positive value 
in the tail region. If the positive imaginary part is cut, 
it increases the reaction cross section by 2~\% for a $^{12}$C 
projectile and by 1~\% for Ne isotopes. 
This cut is used in this paper. If the cut is not taken, $F$ becomes 1.0 
and hence the resultant reaction cross sections for Ne isotopes are 
increased by 1~\% from the present results. 
This numerical ambiguity does not change the conclusion of this paper, 
since the ambiguity is tiny.

\subsection{AMD analysis for Ne isotopes}
\label{AMD analysis for Ne isotopes}

Table~\ref{tab:AMD-result} shows AMD results for
the ground-state properties of Ne isotopes, i.e., the spin-parity
($J^{\pi}$), the one-neutron separation energy $S_{\rm-1 n}$
and the values of deformation parameters $(\bar{\beta},\bar{\gamma})$.
The AMD yields the same $J^{\pi}$ as the data displayed on
the web site~\cite{ICN},
although they are not established experimentally for $^{27,29,31}$Ne.
Particularly for $^{31}$Ne in the IOI region,
the ground state has $J^{\pi}=3/2^{-}$ and small $S_{\rm-1 n}$ consistent
with the data 0.290 $\pm$1.640~\cite{Jurado}.
For $^{28}$Ne corresponding to the boundary of
the IOI region, the main component of the
ground state is the oblate state with $\bar{\beta}=0.28$, but
it is strongly mixed by the prolate state with $\bar{\beta}=0.5$.
The deformation parameter $\bar{\beta}$ decreases as $A$ increases 
from 20 to 25 
and increases as $A$ increases from 25 to 32. The deformation becomes
smallest at $A=25$.

\begin{table}
\caption{ Ground-state properties of Ne isotopes predicted by the AMD.
For $^{28}$Ne, the oblate state with $\bar{\beta}=0.28$ is 
the main component of
the ground state, but it is strongly mixed by
the prolate state with $\bar{\beta}=0.5$.
}
\label{tab:AMD-result}
\begin{center}
\begin{tabular}{cccccc}\hline \hline
 nuclide   & $J^{\pi}$(exp) & $J^{\pi}$(AMD) & $S_{\rm -1n}$ [MeV]
 &  $\bar{\beta}$ & $\bar{\gamma}$  \\ \hline
 $^{20}$Ne & 0$^+$     & 0$^+$     &       & 0.46       &  0$^{^\circ}$  \\
 $^{21}$Ne & 3/2$^+$   & 3/2$^+$   & 7.111 & 0.44       &  0$^{^\circ}$  \\
 $^{22}$Ne & 0$^+$     & 0$^+$     & 9.779 & 0.39       &  0$^{^\circ}$  \\
 $^{23}$Ne & 5/2$^+$   & 5/2$^+$   & 6.021 & 0.32       &  0$^{^\circ}$  \\
 $^{24}$Ne & 0$^+$     & 0$^+$     & 8.231 & 0.25       & 60$^{^\circ}$  \\
 $^{25}$Ne & 1/2$^+$   & 1/2$^+$   & 4.339     & 0.20
           & 31$^{^\circ}$    \\
 $^{26}$Ne & 0$^+$     & 0$^+$     & 5.153 & 0.22       & 0.1$^{^\circ}$ \\
 $^{27}$Ne & (3/2$^+$) & 3/2$^+$   & 1.767      & 0.27
           & 13.6$^{^\circ}$  \\
 $^{28}$Ne & 0$^+$     & 0$^+$     & 3.123 & 0.28(0.50) &  0$^{^\circ}$  \\
 $^{29}$Ne & (3/2$^+$) & 1/2$^+$   & 1.321 & 0.43       &  0$^{^\circ}$  \\
 $^{30}$Ne & 0$^+$     & 0$^+$     & 2.025 & 0.39       &  0$^{^\circ}$  \\
 $^{31}$Ne &           & 3/2$^-$   & 0.248 & 0.41       &  0$^{^\circ}$  \\
 $^{32}$Ne & 0$^+$     & 0$^+$     & 1.012 & 0.33       &  0$^{^\circ}$  \\
 \hline
\end{tabular}
\end{center}
\end{table}

\begin{figure}[htbp]
\begin{center}
\hspace*{-0.8cm}
 \includegraphics[width=0.36\textwidth,clip]{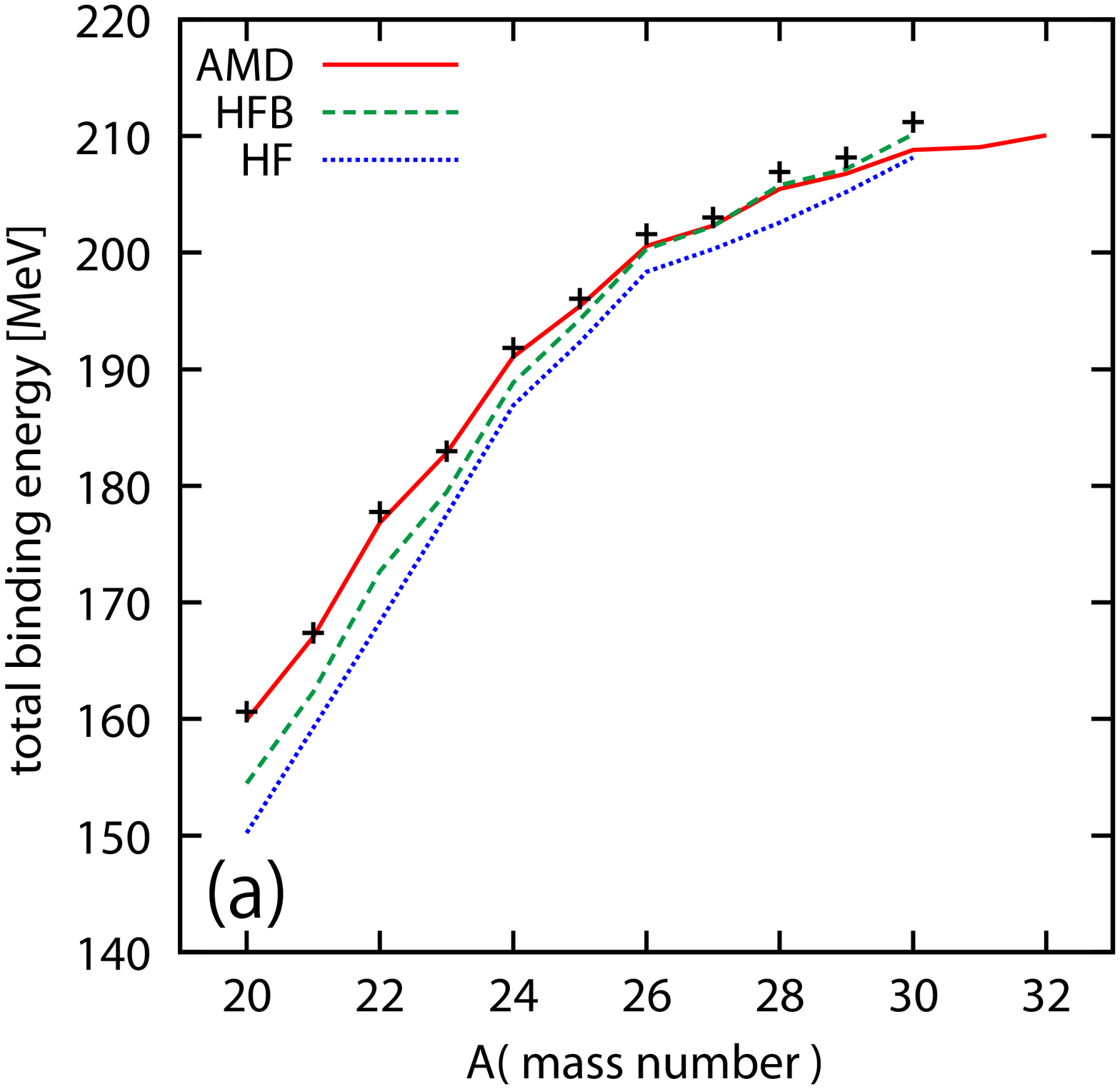}
\hspace*{-0.5cm}
 \includegraphics[width=0.35\textwidth,clip]{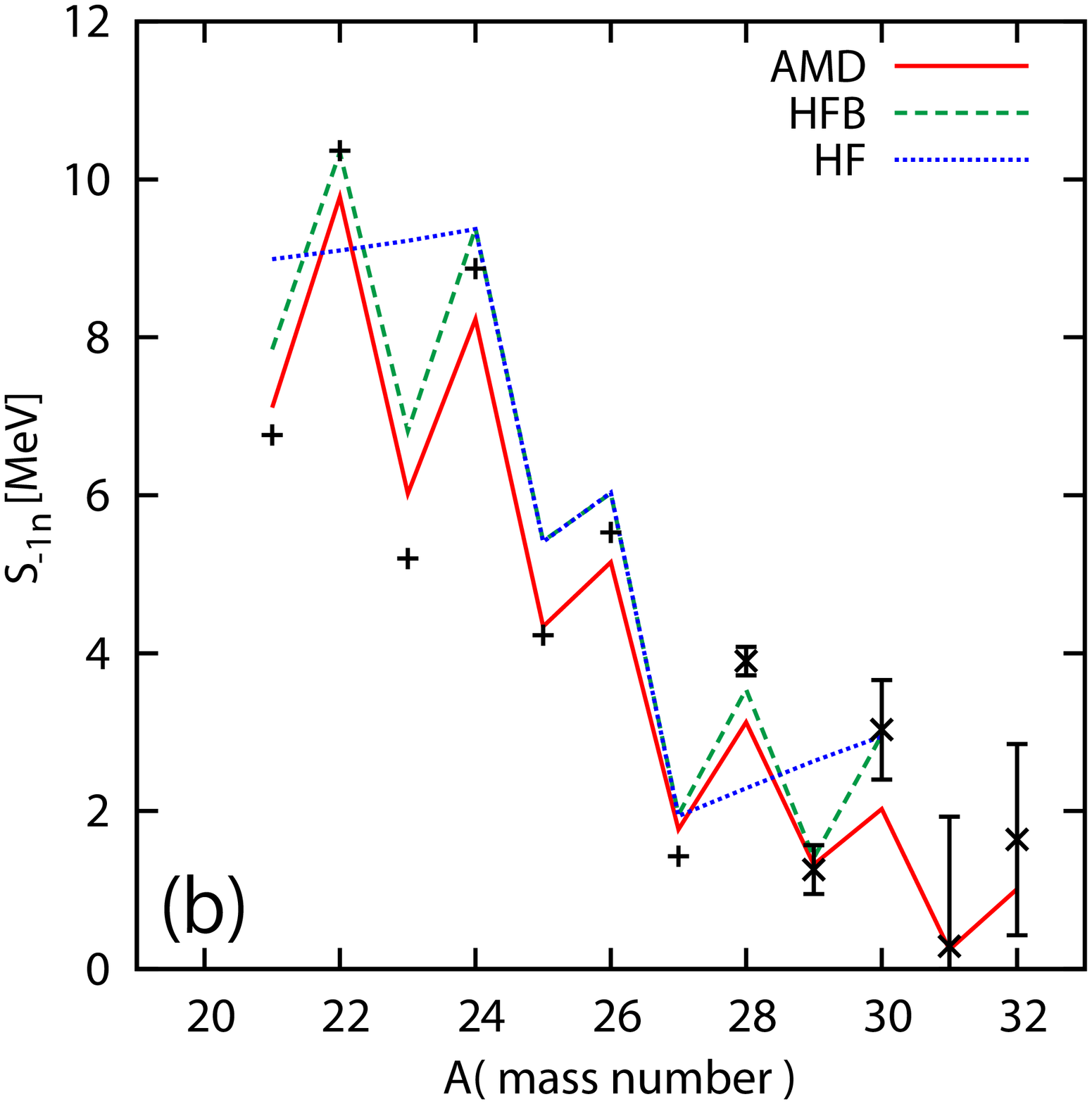}
 \caption{(Color online)
 Results of the AMD, the spherical Gogny-HF and the 
 spherical Gogny-HFB calculation 
 for (a) the total binding energy and (b) the
 one-neutron separation energy of Ne isotopes.
 The dotted, dashed and solid lines represent results of Gogny-HF, Gogny-HFB
and AMD calculations.
  In the spherical HF calculations, the nuclei with $A > 30$ are unbound.
  The experimental data are taken from Refs.~\cite{AusiWapstra,Jurado,ICN}.
  }
 \label{Fig-AMD-Sn}
\end{center}
\end{figure}

Figure~\ref{Fig-AMD-Sn} plots (a) the total binding energy and
(b) $S_{\rm -1n}$ as a function of $A$; here the data are taken from
Refs.~\cite{AusiWapstra,Jurado}.
In the HF and HFB calculations,
the spherical shape is imposed with the filling approximation, and
the nucleus with $A > 30$ are unbound.
The Gogny-HF calculations (dotted lines) underestimate
the total binding energy systematically. 
This situation is improved by the Gogny-HFB calculations
(dashed lines).
The Gogny-AMD calculations (solid lines) yield even better agreement
with the data.
For $S_{\rm -1n}$, the Gogny-HF calculations can not reproduce
the even-odd difference well, but this problem is improved by
the Gogny-HFB calculations. Thus the pairing effect is important
for $S_{\rm -1n}$. The Gogny-AMD calculations almost reproduce
the even-odd difference for all Ne isotopes from A=21 to 32,
but slightly underestimate the experimental even-odd difference.
This may indicate that the pairing effect is partly included in the
Gogny-AMD calculations. The deformation parameter $\bar{\beta}$ 
is 0.33 for $^{32}$Ne and 0.41 for $^{31}$Ne. The reduction of 
$\bar{\beta}$ from 0.41 to 0.33 may come partly from the pairing effect.

\begin{figure}[htbp]
\begin{center}
 \includegraphics[width=0.35\textwidth,clip]{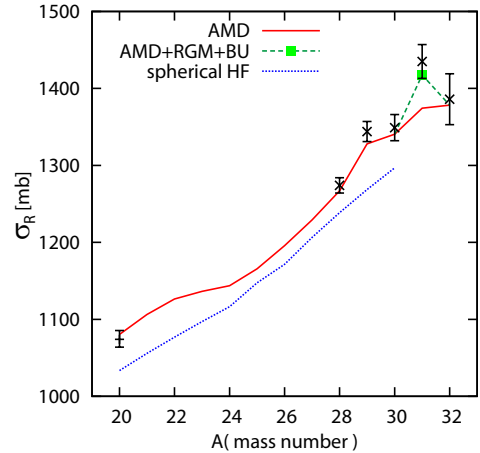}
 \caption{(Color online) Reaction cross sections for scattering of Ne
isotopes on
$^{12}$C at 240~MeV/nucleon.
 The solid (dotted) line represents results of the AMD (spherical Gogny-HF)
 calculations.
 The dashed line with a closed square is the AMD calculation with
 the tail and breakup corrections.
 The experimental data for $A=28-32$
 are taken from Ref.~\cite{Takechi}.
 The data for $^{20}$Ne is deduced
 from measured $\sigma_{\rm I}$ at around 1~GeV/nucleon~\cite{Ne20-sigmaI}
 with the Glauber model~\cite{Takechi}.
  }
 \label{Fig-AMD-HF-sigma-R}
\end{center}
\end{figure}

Figure~\ref{Fig-AMD-HF-sigma-R} represents $\sigma_{\rm R}$
for scattering of Ne isotopes on $^{12}$C at 240~MeV/nucleon.
The AMD calculations (solid line) succeed in reproducing
the data~\cite{Takechi}, while
the spherical Gogny-HF calculation (dotted line) undershoots the data;
note that the spherical Gogny-HFB calculation yields the same result as
the spherical Gogny-HF calculation within the thickness of line.
The nuclei with $A > 30$ are unbound in these spherical calculations.
The enhancement from the dotted line to the solid line comes from
the deformation of the ground state, since the deformation is a main
difference between the two calculations.
The AMD results are consistent with all the data except $^{31}$Ne.
The underestimation of the AMD result for $^{31}$Ne comes from
the inaccuracy of the AMD density in its tail region.

The tail problem is solved by the following 
resonating group method~\cite{Minomo:2011bb}.
In principle the ground state 
$\Phi(^{31}{\rm Ne}; 3/2^-_1)$ 
of $^{31}$Ne can be expanded in terms of
the ground and excited states $\Phi(^{30}{\rm Ne}; J^\pi_n)$ 
of $^{30}$Ne. This means that
the ground state of $^{31}$Ne is described by
the $^{30}$Ne+n cluster model with the core ($^{30}$Ne) excitations.
The cluster-model calculation can be done with the resonating
group method (RGM) in which the ground and excited states
of $^{30}$Ne are constructed by the AMD:
\begin{align}
 \Phi(^{31}{\rm Ne}; 3/2^-_1) = 
  \sum_{nJ\pi}{\cal A}
  \left\{
   \chi_{nlj}^{}(r) Y_{lm}^{}(\hat{\vrr})
    \Phi(^{30}{\rm Ne}; J^\pi_n)\phi_{n}^{}
  \right\},
\end{align}
where $\phi_{n}^{}$ is the intrinsic wave function of last neutron and  
$\chi_{nlj}^{}$ is the relative wave function  between the last neutron 
and the core ($^{30}$Ne).  
Here the wave function of $^{30}$Ne includes many excited states 
with positive- and negative-parity below 10 MeV in excitation energy. 
This AMD+RGM calculation is quite time consuming, but it was done for
$^{31}$Ne. The tail correction to $\sigma_{\rm R}$ is 35~mb.

\begin{figure}[htbp]
\begin{center}
\hspace*{-0.4cm}
 \includegraphics[width=0.50\textwidth,clip]{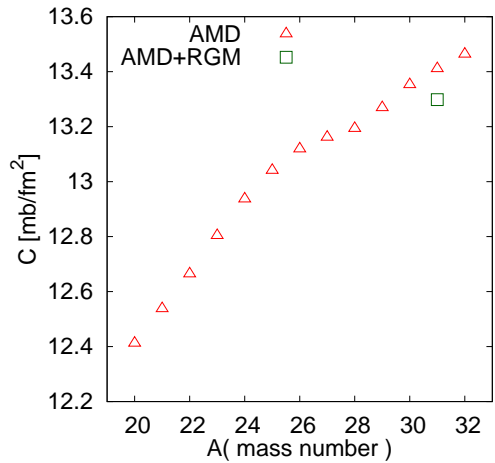}
 \caption{(Color online) $A$ dependence of the coefficient $C$. 
Open triangles show results of the AMD calculations, whereas an open 
square corresponds to a result of the AMD+RGM calculation for $^{31}$Ne. 
There is a non-negligible difference between the AMD and the AMD+RGM result 
for $^{31}$Ne.  
Thus, $C$ is slightly reduced, when the density has the halo structure. 
 }
 \label{Fig-C-A}
\end{center}
\end{figure}

\begin{figure}[htbp]
\begin{center}
 \includegraphics[width=0.35\textwidth,clip]{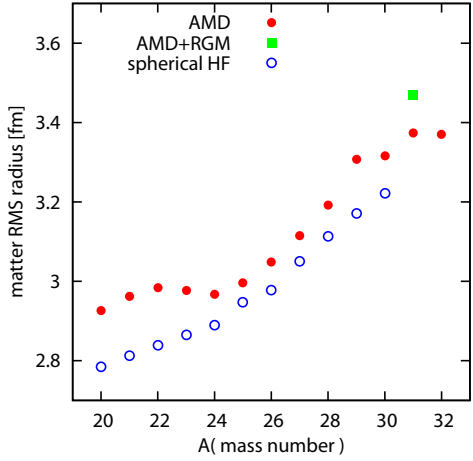}
 \caption{(Color online) Matter RMS radii of Ne isotopes calculated with
 the AMD, the AMD+RGM, and the spherical HF model.
 The closed circle represents results of the AMD, and the closed square
 denotes a result of the AMD+RGM model for $^{31}$Ne.
 The opened circles are results of the spherical HF calculation.
 }
 \label{Fig-AMD-HF-RMS}
\end{center}
\end{figure}

\begin{figure}[htbp]
\begin{center}
 \includegraphics[width=0.35\textwidth,clip]{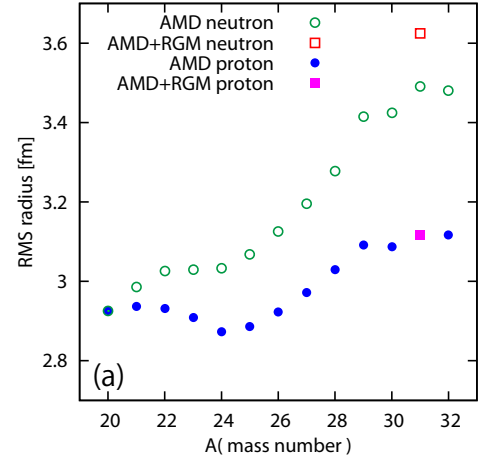}
 \includegraphics[width=0.35\textwidth,clip]{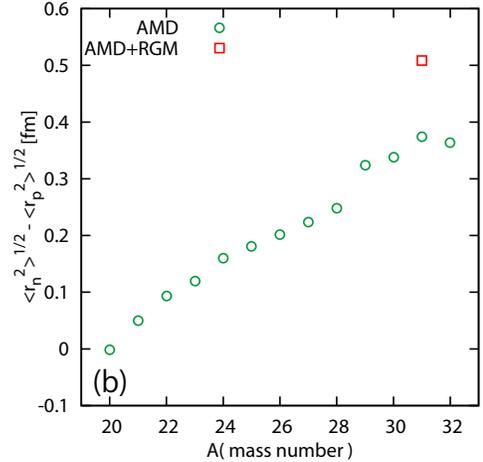}
 \caption{(Color online) Neutron and proton RMS radii of Ne isotopes
 calculated with  the AMD and the AMD+RGM model.
 In panel (a),
 the closed (opened) circle represents the proton (neutron) RMS radius
 calculated with the AMD and the closed (opened) square denotes a
 result of the AMD+RGM calculation for proton (neutron) of $^{31}$Ne;
 note that the AMD+RGM result agrees with the AMD result for proton
 RMS radius.
 Panel (b) shows the difference between
 neutron and proton RMS radii. The open circle (square) stands
 for the AMD (AMD+RGM) result.
  }
 \label{Fig-AMD-HF-RMS_pn}
\end{center}
\end{figure}

\begin{figure*}[htbp]
\begin{center} 
 \includegraphics[width=0.55\textwidth,clip]{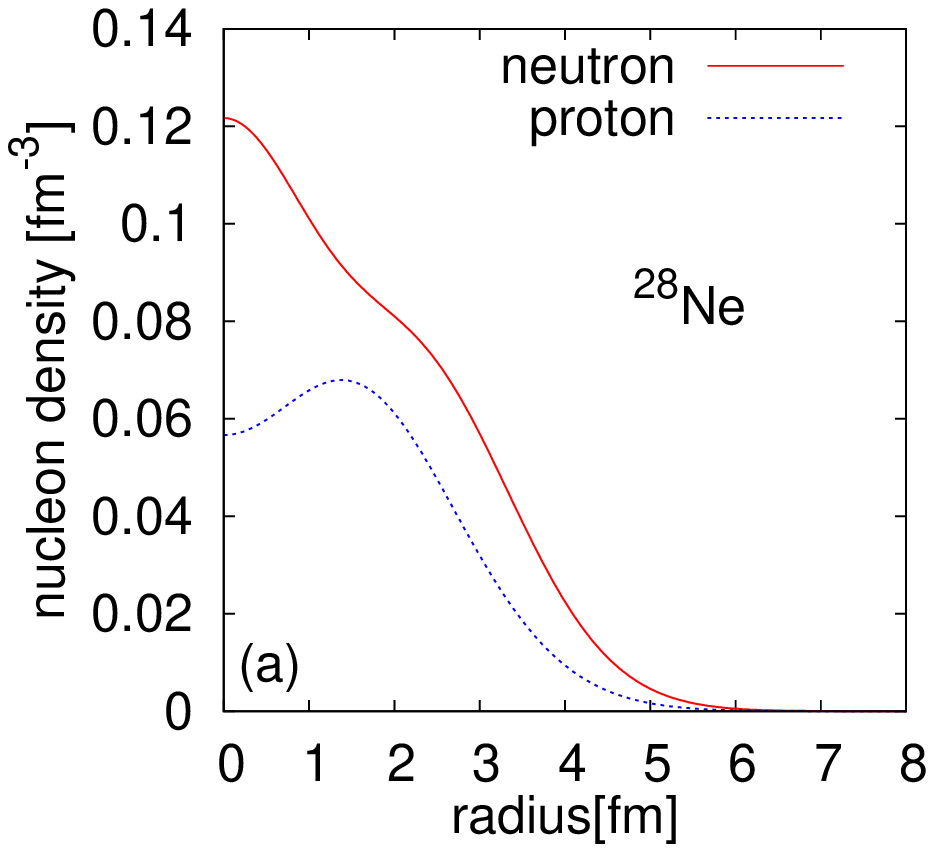}
\hspace*{-2.2cm}
 \includegraphics[width=0.55\textwidth,clip]{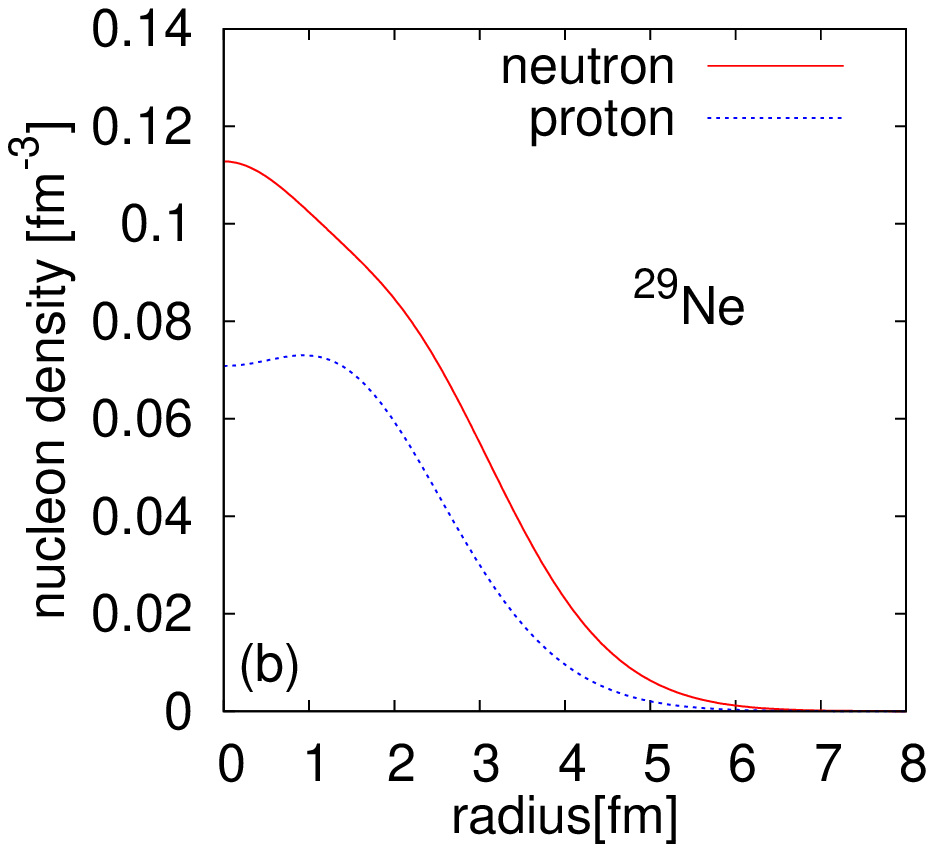} \\
 \includegraphics[width=0.55\textwidth,clip]{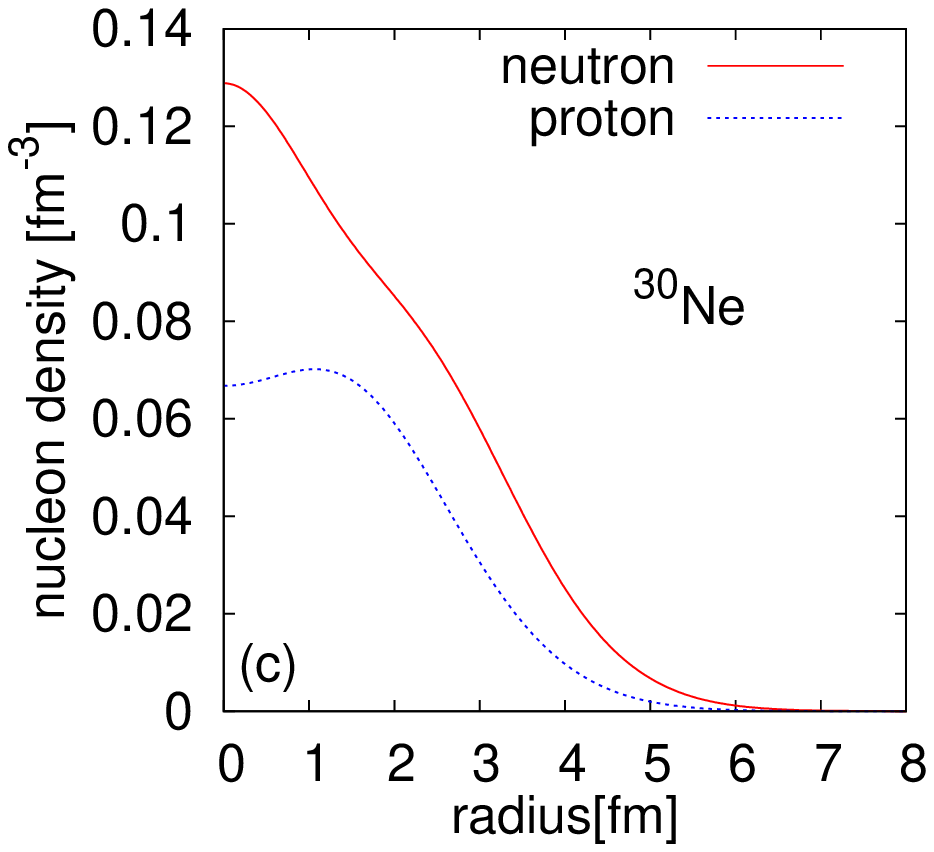} 
\hspace*{-2.2cm}
 \includegraphics[width=0.55\textwidth,clip]{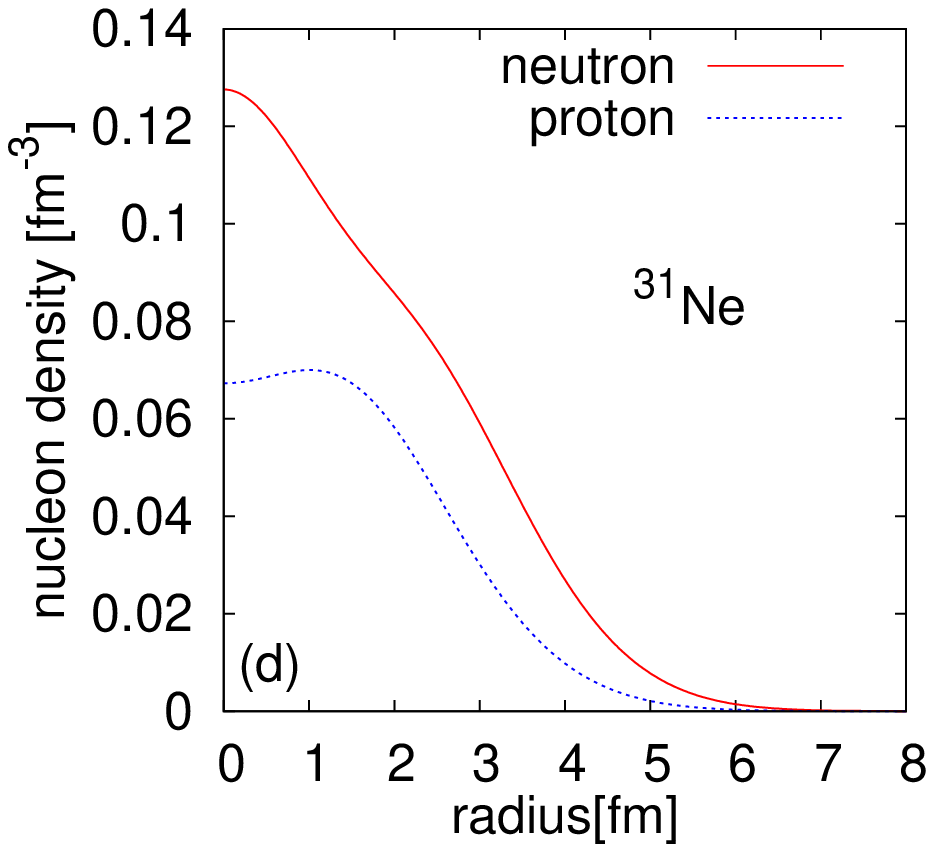} \\
 \includegraphics[width=0.55\textwidth,clip]{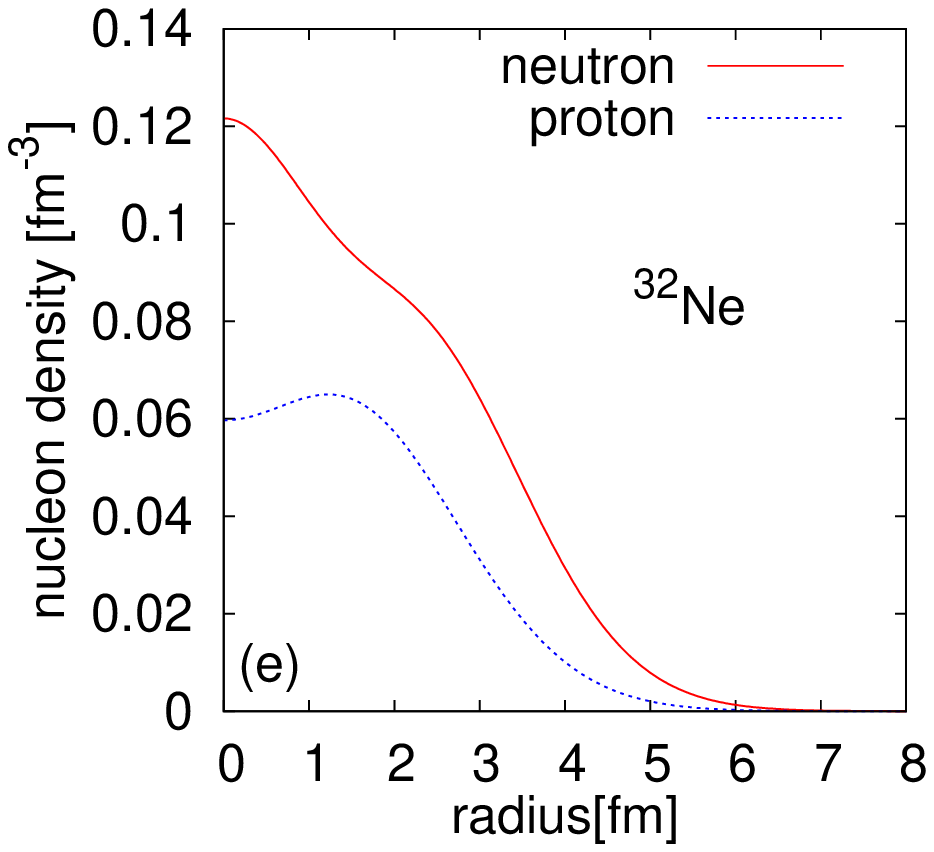}
\hspace*{-2.2cm}
 \includegraphics[width=0.55\textwidth,clip]{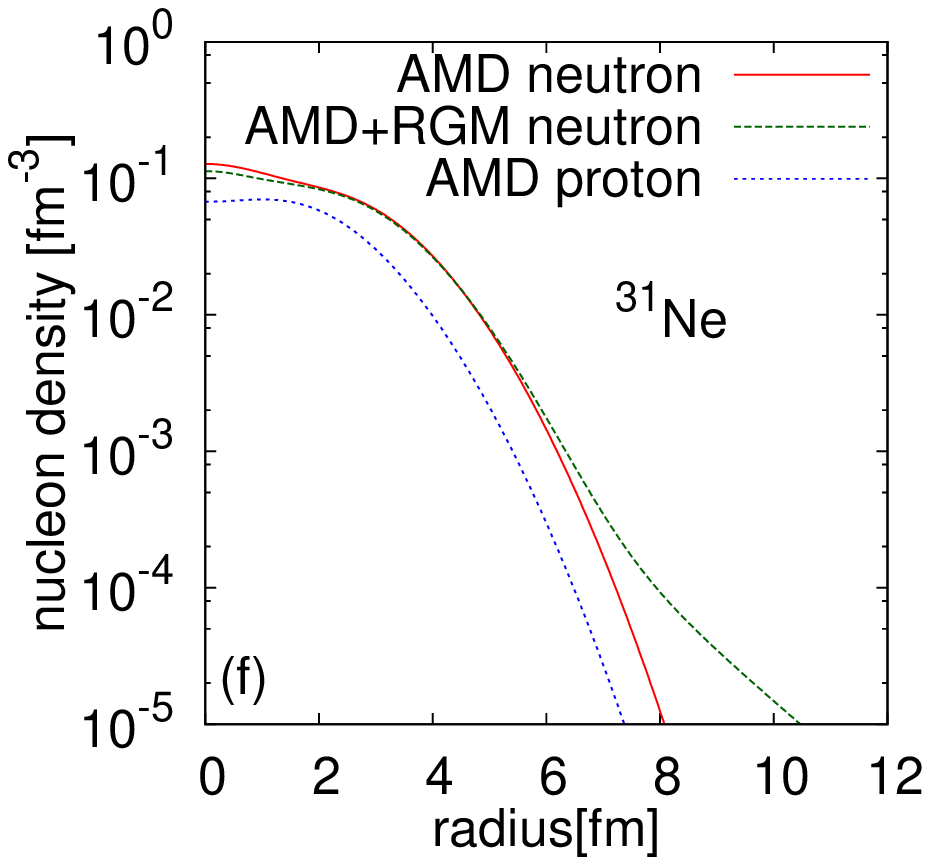}
 \caption{(Color online) Neutron and proton density profiles for
 (a) $^{28}$Ne, (b) $^{29}$Ne,
 (c) $^{30}$Ne, (d) $^{31}$Ne and (e) $^{32}$Ne on a linear scale
 and (f) $^{31}$Ne  on a logarithmic scale.
 In panels (a)-(f), the solid (dotted) line presents
 the neutron (proton) density profile calculated with the AMD method,
 whereas the dashed line in panel (f) is the neutron density profile
 calculated with the AMD+RGM method.
   }
 \label{Fig-AMD-density-pn}
\end{center}
\end{figure*}

For a weakly bound system such as $^{31}$Ne, furthermore,
the projectile breakup effect is not perfectly negligible.
This effect is simply estimated
by assuming the potential model for the $^{30}$Ne+n system and
solving the three-body dynamics of the $^{30}$Ne+n+$^{12}$C system
with the method of continuum discretized coupled channels
(CDCC)~\cite{CDCC-review1,CDCC-review2}.
CDCC is an accurate method for treating exclusive reactions such as
elastic scattering and elastic-breakup reactions.
The theoretical foundation of CDCC is shown
in Refs.~\cite{CDCC-foundation1,CDCC-foundation2,CDCC-foundation3}.
CDCC succeeded in reproducing data on the scattering of
stable and unstable projectiles~\cite{CDCC-review1,CDCC-review2,
Rusek,Tostevin2,Davids,Mortimer,Eikonal-CDCC,Matsumoto3,Howell,Rusek2,
Matsumoto4,Moro,THO-CDCC,4body-CDCC-bin,Matsumoto:2010mi,Avrigeanu}.
Here the interactions between $^{30}$Ne and $^{12}$C
and between n and $^{12}$C are constructed with the DFM with 
the Melbourne $g$-matrix,
and the potential between $^{30}$Ne and n is made
with the well-depth method; see Refs.~\cite{ERT,30Ne}
for the potential parameters.
The correction is 10~mb corresponding to 0.7\% of $\sigma_{\rm R}$. 
In Fig.~\ref{Fig-AMD-HF-sigma-R}, the dashed line stands for the 
AMD result with the tail and breakup corrections for $^{31}$Ne. The result is
consistent with the data for $^{28-32}$Ne.

The reaction cross section $\sigma_{\rm R}$ is sensitive to the RMS radii,
$\sqrt{\langle \vrr^2 \rangle_{\rm P}}$ and
$\sqrt{\langle \vrr^2 \rangle_{\rm T}}$,
of projectile and target. Actually,
the DFM calculation for $^{20-32}$Ne projectiles shows that
\bea
\sigma_{\rm R}=C \pi \left[
{\sqrt {\langle \vrr^2 \rangle_{\rm P}}} +
{\sqrt {\langle \vrr^2 \rangle_{\rm T}}}
\right]^2, 
\label{Sigma-RMS}
\eea
where $C$ is a slowly varying function of $A$ around 
$C=12.4 \sim 13.5$~[mb/fm${^2}$], as shown in Fig.~\ref{Fig-C-A}. 
Now it is assumed that the projectile density has a deformed well shape.
If the volume conservation is imposed
with the general shape in Eq.~(\ref{eq:surf}),
the matter squared radius $\langle \vrr^2 \rangle_{\rm P}$ of projectile
is described by
\begin{align}
 \langle \vrr^2 \rangle_{\rm P} &= \langle \vrr^2 \rangle_{0}
 \biggl[ 1 + \frac{5}{4\pi} \sum_{\lambda\mu}|\alpha_{\lambda\mu}|^2 \biggr] \\
 &=\langle \vrr^2 \rangle_{0}
 \biggl[ 1 + \frac{5}{4\pi} (\beta_2^2+\beta_4^2 + \cdots) \biggr]
\label{box_RMS_2}
\end{align}
up to the second order in the deformation parameters $\{\alpha_{\lambda\mu}\}$,
where $\langle \vrr^2 \rangle_{0}$ is the matter squared radius in
the spherical limit.
The triaxial parameter $\gamma$ 
does not appear in Eq.~\eqref{box_RMS_2}.
This means that the triaxial deformation little affect
the RMS radius and then $\sigma_{\rm R}$;
it has been confirmed by utilizing the Woods-Saxon model in place of 
the AMD model,
in which the triaxial parameter $\bar{\gamma}$ is optimized and
cannot be changed artificially.
For $^{27}$Ne, we varied the $\gamma$ parameter
from $0^{\circ}$ to $60^{\circ}$ with fixing $\beta_2=0.273$ ($\beta_4=0$)
corresponding to the $\bar{\beta}=0.27$ predicted by the AMD model,
but the resultant reaction cross section
changes only by 0.2\%.  Figure~\ref{Fig-AMD-HF-RMS} shows the RMS radii
of the spherical-HF, AMD, AMD+RGM calculations.
The differences among three calculations for the RMS radii are similar to
those for $\sigma_{\rm R}$ shown in Fig. \ref{Fig-AMD-HF-sigma-R}.
The difference between the AMD and AMD+RGM calculations
for the RMS radius of $^{31}$Ne is appreciable, indicating that
the tail correction is significant for this very-weakly bound system.

Finally we compare the neutron RMS radius
$\sqrt{\langle \vrr_n^2 \rangle}$ with
the proton one $\sqrt{\langle \vrr_p^2 \rangle}$ in order to
see the isovector components of the Ne-isotope densities.
Figure~\ref{Fig-AMD-HF-RMS_pn} shows
$A$-dependence of $\sqrt{\langle \vrr_n^2 \rangle}$ and
$\sqrt{\langle \vrr_p^2 \rangle}$ for Ne isotopes.
In panel (a), the neutron and proton RMS radii increase with $A$,
when $A \ge 24$. For $A=20-24$, the proton RMS radii have a bump.
This implies that
at $A=20-22$ the proton-neutron correlation is strong and hence
the alpha-clustering grows.
Panel (b) shows the difference
$\sqrt{\langle \vrr_n^2 \rangle}
-\sqrt{\langle \vrr_p^2 \rangle}$ as a function of $A$.
The difference also goes up as $A$ increases.
There is a sizable jump between $A=28$ and 29,
since the deformation $\bar{\beta}$
is around 0.25 at $A=24-28$ but around 0.4 at $A=29-32$.
As a result of this gap, the radius difference is around 0.35~fm,
indicating
that Ne isotopes are either skin or skin-like nuclei for $A=29, 30, 32$.
For $^{31}$Ne, the radius difference calculated with the AMD+RGM method
is about 0.5~fm that is significantly bigger than 0.35~fm. This implies that
$^{31}$Ne is a halo nucleus.
These interpretations are more obvious through
the neutron and proton density profiles shown in
Fig.~\ref{Fig-AMD-density-pn}. Panels (a), (b), (c), (d), and (e) show 
the density profiles for $^{28-32}$Ne, respectively. 
$^{29,30,32}$Ne have the neutron-skin structure. In panel (f), 
the density profiles for $^{31}$Ne is plotted 
on a logarithmic scale. The neutron density (dashed line)
calculated with the AMD+RGM method has a long-range tail,
indicating that $^{31}$Ne has the halo structure.

\subsection{Woods-Saxon mean-field model}
\label{Sec:DWS}

\begin{figure}[htbp]
\begin{center}
\hspace*{-0.5cm}
 \includegraphics[width=0.4\textwidth,clip]{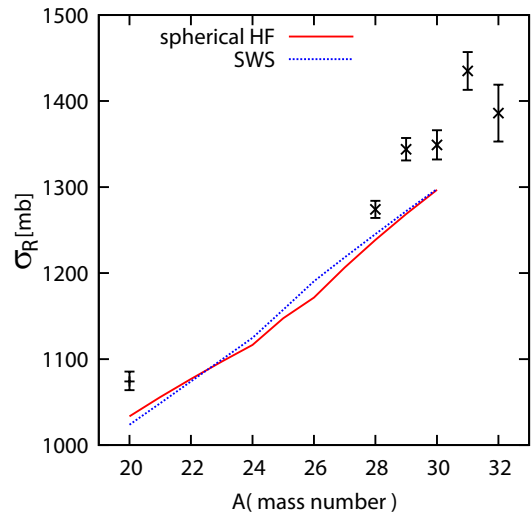}
 \caption{(Color online)
Reaction cross sections
calculated for Ne isotopes calculated 
with the SWS model and the Gogny-HF method.
The spherical shape is imposed for both the calculations.
The dotted line represents results of the SWS model, while
the solid line corresponds to the spherical Gogny-HF results.
The nucleus with $A > 30$ are unbound.
The experimental data are taken from Ref.~\cite{Takechi}.
 }
 \label{Fig-HF-SWSx}
\end{center}
\end{figure}

In this subsection, the results of the Woods-Saxon mean-field model 
are investigated.  First of all,
in order to see that the parameter set of the Woods-Saxon potential
is reasonable, the spherical case is studied.
Figure \ref{Fig-HF-SWSx} shows the reaction cross sections
for Ne isotopes calculated with the spherical Woods-Saxon (SWS) model
(neglecting the pairing correlation) and the spherical Gogny-HF method.
The SWS model (dotted line) well simulates
results of the spherical Gogny-HF calculation (solid line).
This means that the SWS model yields almost the same matter radius
as the spherical Gogny-HF calculation.
The SWS model with the present parameter set proposed by
R.~Wyss~\cite{WyssPriv} is thus a handy way
of simulating the spherical Gogny-HF calculation.

\begin{figure}[htbp]
\begin{center}
 \includegraphics[width=0.35\textwidth,clip]{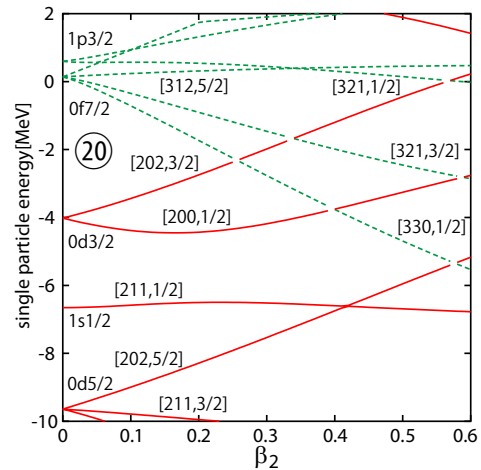}
 \caption{(Color online) The neutron Nilsson diagram for $^{30}$Ne
 in the deformed WS model, where the other parameters are fixed
 to $\beta_4=\gamma=0$. The solid (dashed) lines correspond to 
 the positive (negative) parity orbits. The Nilsson asymptotic quantum 
 numbers [$N$,$n_3$,$\Lambda$,$\Omega$] are attached. The number 20 stands 
 for a neutron magic number in the spherical limit. 
 }
 \label{Nilsson-diagram}
\end{center}
\end{figure}

Figure~\ref{Nilsson-diagram} shows the neutron Nilsson diagram
for $^{30}$Ne calculated with the DWS model. 
It is emphasized that the relatively large shell gap with 
$N=20$ is observed at the spherical shape ($\beta_2=0$).
Comparing this figure with the Nilsson diagram calculated with 
the AMD model in Fig. 2 of Ref.~\cite{Kimura},
one can see that the Nilsson diagram of the DWS model is close
to that of the AMD model.
In both the models (AMD and DWS), 
the [2,0,0,1/2] and the [3,3,0,1/2] orbit 
in terms of the Nilsson asymptotic quantum numbers 
[$N$,$n_3$,$\Lambda$,$\Omega$]
cross each other at $\beta_2 \approx 0.4$, although
the single-particle energy at the crossing point is
$-3$~MeV in the AMD model and $-4$~MeV in the DWS model.
It is well-known that the occupation of this down-sloping orbit [3,3,0,1/2]
derives the system to deform near $N\approx 20$.

Next, the results of the deformed Woods-Saxon (DWS) model
with deformation obtained by the microscopic-macroscopic
(Strutinsky) method are discussed.
The pairing correlation is included within the Kruppa-BCS approximation,
which strongly affects the resultant deformations.
Two cases of the pairing strengths determined by
the smooth pairing gap $\widetilde\Delta=12/\sqrt{A}$ or $4/\sqrt{A}$~MeV are considered;
$\widetilde\Delta=12/\sqrt{A}$~MeV is often used as a typical value
for medium-heavy nuclei,
while $\widetilde\Delta=4/\sqrt{A}$~MeV is used to simulate the weak pairing case.
The smooth pairing gap $\widetilde\Delta$ is usually supposed to correspond
to the even-odd mass difference.  
However, in the light mass nuclei,
like the Ne isotopes considered in the present work,
the even-odd mass differences contain considerable amount
of the shell effect of deformed mean-field,
and the weaker pairing correlations are suggested~\cite{SDN98}.
The small value $\widetilde\Delta=4/\sqrt{A}$~MeV is chosen in accordance 
with it.
In Tables~\ref{tab:WSS_P12-beta} and~\ref{tab:WSS_P04-beta},
the results of the deformation parameters ($\beta_2,\beta_4$) and
the neutron and proton pairing gaps $\Delta_{\rm n}$ and $\Delta_{\rm p}$
are summarized; all nuclei are calculated to be axially symmetric
in their ground states.
If the standard pairing is used, all even-even isotopes
turn out to be spherical, which contradicts the result
of the AMD calculation.
Even with weaker pairing, $^{25,26,28,30}$Ne are calculated to be spherical;
this is because of the relatively large $N=20$ shell gap
(see Fig.~\ref{Nilsson-diagram}), which is in contrast to the prediction
using the tensor force in Ref.~\cite{OMA06}.

\begin{table}
\caption{ Deformation parameters ($\beta_2,\beta_4$)
and the pairing gaps $\Delta_{\rm n}$ and $\Delta_{\rm p}$ obtained
by the Strutinsky calculation with $\widetilde\Delta=12/\sqrt{A}$~MeV.
The nucleus $^{31}$Ne is unbound and $^*$ is attached.
}
\label{tab:WSS_P12-beta}
\begin{center}
\begin{tabular}{ccccc} \hline \hline
 nuclide  & $\beta_2$ & $\beta_4$ &
 $\Delta_{\rm n}$ [MeV] & $\Delta_{\rm p}$ [MeV]  \\ \hline
 $^{20}$Ne & \hspace*{0.25cm}0.000  & \hspace*{0.25cm}0.000   & 2.897 & 2.848 \\
 $^{21}$Ne & \hspace*{0.25cm}0.310  & \hspace*{0.25cm}0.048   & 0.000 & 2.290 \\
 $^{22}$Ne & \hspace*{0.25cm}0.000  & \hspace*{0.25cm}0.000   & 2.801 & 2.826 \\
 $^{23}$Ne & \hspace*{0.25cm}0.200  & \hspace*{0.25cm}0.011   & 0.889 & 2.596 \\
 $^{24}$Ne & \hspace*{0.25cm}0.000  & \hspace*{0.25cm}0.000   & 2.337 & 2.865 \\
 $^{25}$Ne & \hspace*{0.25cm}0.000  & \hspace*{0.25cm}0.000   & 0.000 & 2.849 \\
 $^{26}$Ne & \hspace*{0.25cm}0.000  & \hspace*{0.25cm}0.000   & 2.157 & 2.826 \\
 $^{27}$Ne & $-$0.065            & $-$0.002             & 0.989 & 2.783 \\
 $^{28}$Ne & \hspace*{0.25cm}0.000  & \hspace*{0.25cm}0.000   & 2.345 & 2.783 \\
 $^{29}$Ne & $-$0.048            & $-$0.001             & 1.391 & 2.748 \\
 $^{30}$Ne & \hspace*{0.25cm}0.000  & \hspace*{0.25cm}0.000   & 2.286 & 2.734 \\
$^{31}$Ne&\hspace*{0.40cm}0.101$^*$&\hspace*{0.40cm}0.022$^*$ & 1.618 & 2.623 \\
 $^{32}$Ne & \hspace*{0.25cm}0.000  & \hspace*{0.25cm}0.000   & 2.416 & 2.674 \\
 \hline
\end{tabular}
\end{center}
\end{table}

\begin{table}
\caption{ The same as Table~\ref{tab:WSS_P12-beta} but
with $\widetilde\Delta=4/\sqrt{A}$~MeV. }
\label{tab:WSS_P04-beta}
\begin{center}
\begin{tabular}{ccccc} \hline \hline
 nuclide  & $\beta_2$ & $\beta_4$ &
 $\Delta_{\rm n}$ [MeV] & $\Delta_{\rm p}$ [MeV]  \\ \hline
 $^{20}$Ne & \hspace*{0.25cm}0.336  & \hspace*{0.25cm}0.111    & 0.000 & 0.000 \\
 $^{21}$Ne & \hspace*{0.25cm}0.349  & \hspace*{0.25cm}0.079    & 0.000 & 0.000 \\
 $^{22}$Ne & \hspace*{0.25cm}0.362  & \hspace*{0.25cm}0.051    & 0.000 & 0.000 \\
 $^{23}$Ne & \hspace*{0.25cm}0.291  & \hspace*{0.25cm}0.052    & 0.000 & 0.000 \\
 $^{24}$Ne & \hspace*{0.25cm}0.186  & \hspace*{0.25cm}0.018    & 0.000 & 1.434 \\
 $^{25}$Ne & \hspace*{0.25cm}0.000  & \hspace*{0.25cm}0.000    & 0.000 & 1.657 \\
 $^{26}$Ne & \hspace*{0.25cm}0.000  & \hspace*{0.25cm}0.000    & 0.000 & 1.654 \\
 $^{27}$Ne & \hspace*{0.25cm}0.122  & $-$0.001              & 0.000 & 1.471 \\
 $^{28}$Ne & \hspace*{0.25cm}0.001  & \hspace*{0.25cm}0.000    & 0.966 & 1.638 \\
 $^{29}$Ne & \hspace*{0.25cm}0.067  & $-$0.001              & 0.000 & 1.572 \\
 $^{30}$Ne & \hspace*{0.25cm}0.000  & \hspace*{0.25cm}0.000    & 0.000 & 1.611 \\
$^{31}$Ne &\hspace*{0.40cm}0.291$^*$&\hspace*{0.40cm}0.106$^*$ & 0.000 & 0.000 \\
 $^{32}$Ne & \hspace*{0.25cm}0.278  & \hspace*{0.25cm}0.098    & 0.922 & 0.000 \\
 \hline
\end{tabular}
\end{center}
\end{table}

In order to check the consistency of the DWS model,
we compare the obtained deformations with the systematic
Gogny D1S HFB calculations~\cite{HFBwoAMP},
the results of which are available on the web site~\cite{HFBchart}.
Those results are also axially symmetric in Ne isotopes,
and the $(\beta_2,\beta_4)$ deformation parameters are extracted
in such a way that
$\langle r^2Y_{20} \rangle_{\rm uni}/\langle r^2 \rangle_{\rm uni}$ and
$\langle r^4Y_{40} \rangle_{\rm uni}/\langle r^2 \rangle_{\rm uni}^2$
calculated with the uniform density in Eq.~(\ref{eq:unirho})
reproduce the corresponding values of Gogny D1S HFB calculations
tabulated in Ref.~\cite{HFBchart}.
The resultant $(\beta_2,\beta_4)$ are listed in Table~\ref{tab:D1SHFB-beta}.
It can be seen that $(\beta_2,\beta_4)$ calculated with the weaker pairing
in Table~\ref{tab:WSS_P04-beta} well corresponds to those of the deformed
Gogny D1S HFB results.  This result may suggest that
the monopole pairing strength weaker than the standard one
is more appropriate in light nuclei
in accordance with the analysis in Ref.~\cite{SDN98}.

\begin{table}
\caption{ Deformation parameters ($\beta_2,\beta_4$) extracted from
the results of deformed Gogny-D1S HFB calculations of
Refs.~\cite{HFBwoAMP,HFBchart}.
The nucleus $^{31}$Ne is unbound and no data are available.
}
\label{tab:D1SHFB-beta}
\begin{center}
\begin{tabular}{ccc} \hline \hline
 nuclide  & $\beta_2$ & $\beta_4$ \\ \hline
 $^{20}$Ne & \hspace*{0.25cm}0.325  & \hspace*{0.25cm}0.108  \\
 $^{21}$Ne & \hspace*{0.25cm}0.370  & \hspace*{0.25cm}0.085  \\
 $^{22}$Ne & \hspace*{0.25cm}0.355  & \hspace*{0.25cm}0.016  \\
 $^{23}$Ne & \hspace*{0.25cm}0.234  & \hspace*{0.25cm}0.011  \\
 $^{24}$Ne & \hspace*{0.25cm}0.179  & \hspace*{0.25cm}0.011  \\
 $^{25}$Ne & $-$0.047            & \hspace*{0.25cm}0.001  \\
 $^{26}$Ne & $-$0.002            & \hspace*{0.25cm}0.000  \\
 $^{27}$Ne & $-$0.073            & $-$0.005            \\
 $^{28}$Ne & $-$0.006            & \hspace*{0.25cm}0.000  \\
 $^{29}$Ne & $-$0.060            & $-$0.003            \\
 $^{30}$Ne & $-$0.002            & \hspace*{0.25cm}0.000  \\
 $^{31}$Ne & ---               & ---               \\
 $^{32}$Ne & \hspace*{0.25cm}0.246  & \hspace*{0.25cm}0.096  \\
 \hline
\end{tabular}
\end{center}
\end{table}

\begin{figure}[htbp]
\begin{center}
\hspace*{-0.3cm}
 \includegraphics[width=0.355\textwidth,clip]{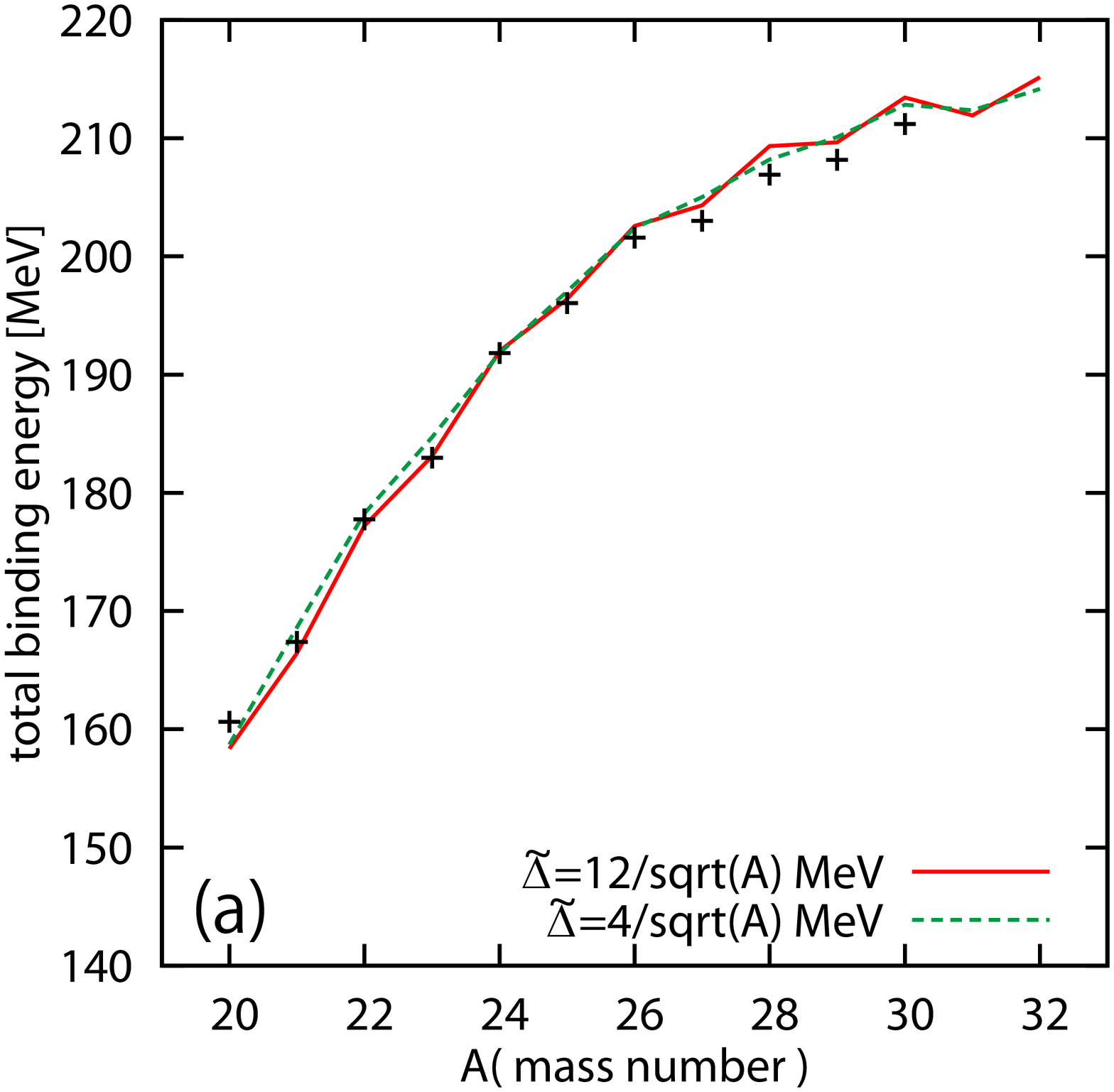}
 \includegraphics[width=0.35\textwidth,clip]{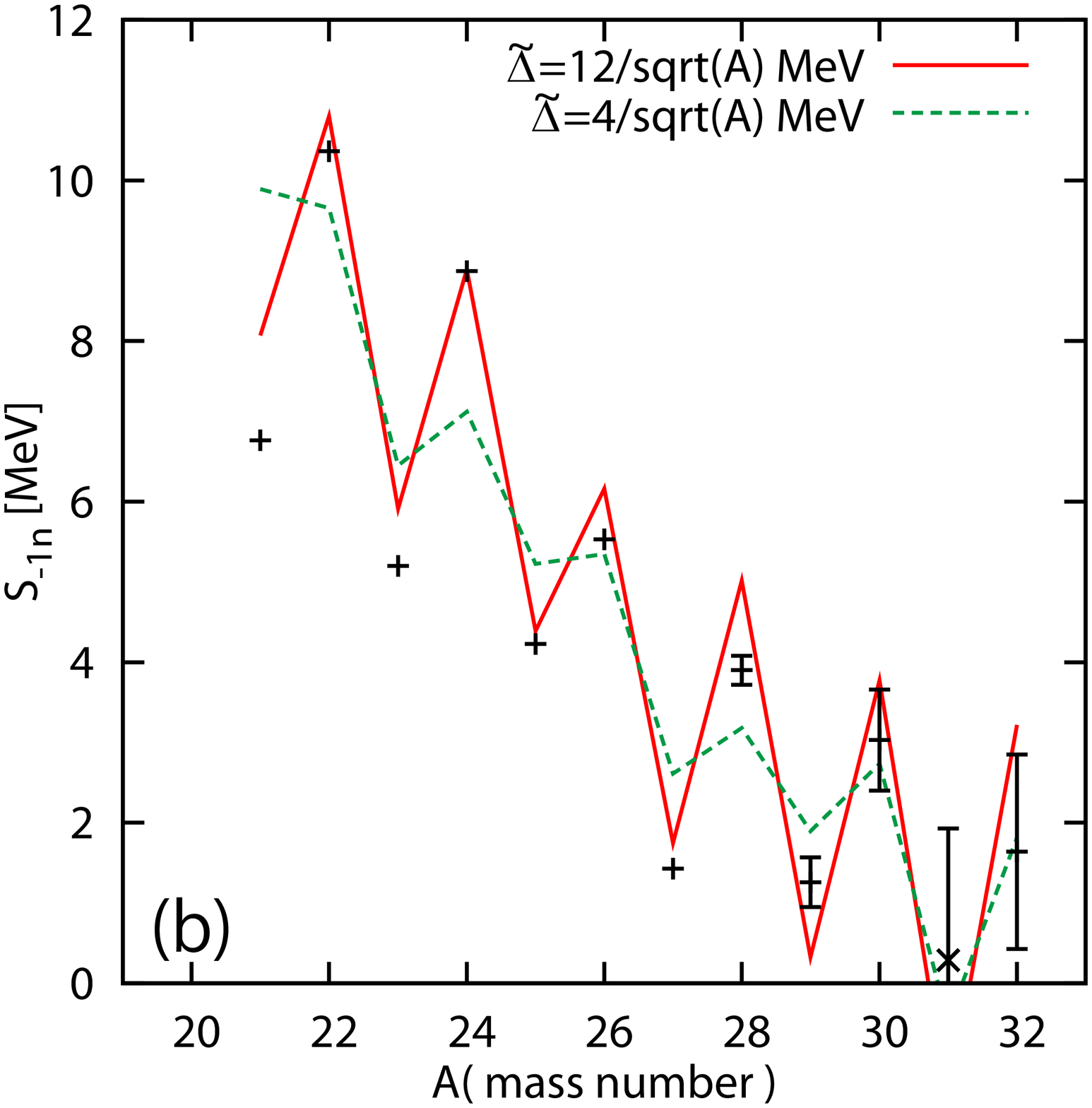}
  \caption{(Color online)
  Results of the DWS model including the pairing correlation
  with $\widetilde\Delta=12/\sqrt{A}$ or $4/\sqrt{A}$~MeV for (a) the total binding energy
  and (b) the one-neutron separation energy $S_{\rm -1n}$ of Ne isotopes.
  The solid line is a result of $\widetilde\Delta=12/\sqrt{A}$~MeV and
  the dashed line of $\widetilde\Delta=4/\sqrt{A}$~MeV.
 }
 \label{Fig:Pairing}
\end{center}
\end{figure}

The calculated binding energies and one-neutron separation energies
in the DWS model are compared with experimental data
in Fig.~\ref{Fig:Pairing}.
The binding energies are slightly overestimated
for unstable isotopes $^{27-30}$Ne, which is mainly due to the liquid drop
energy~\cite{MS67} employed in the present work.
As for the one-neutron separation energies, $S_{\rm -1n}$,
the calculations with the standard pairing strength 
with $\widetilde\Delta=12/\sqrt{A}$ MeV nicely reproduce 
the experimental data, 
although the weaker pairing with $\widetilde\Delta=4/\sqrt{A}$ MeV 
may yield slightly better agreement for $^{30-32}$Ne. 
Similar agreement is seen in the AMD, but 
the predicted deformations of Ne isotopes in the two models
are very different; those in the AMD are larger particularly in
unstable isotopes $^{28-32}$Ne.
This is because the binding energy reflects many structural effects. 
The binding energy is thus  not a good indicator of deformation.

\begin{figure}[htbp]
\begin{center}
 \includegraphics[width=0.38\textwidth,clip]{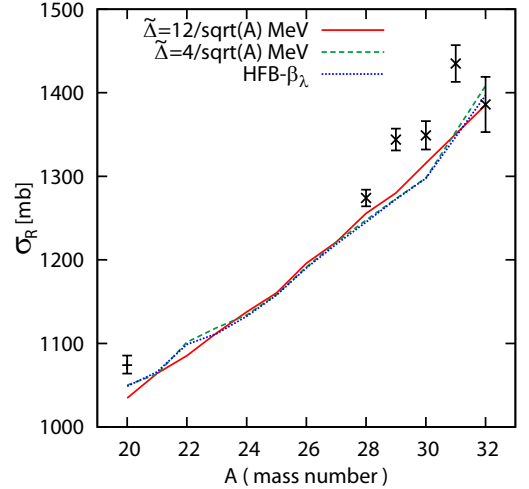}
 \caption{(Color online)
 The reaction cross sections for Ne isotopes calculated
 with the DWS model with some sets of deformation parameters. 
 The solid and dashed lines represent results of the Strutinsky method with
 $\widetilde\Delta=12/\sqrt{A}$ and $4/\sqrt{A}$~MeV. In the dotted line, 
 the deformation is determined by the deformed Gogny-D1S HFB calculation,
 for which the pairing correlation is neglected.
 The experimental data are taken from Ref.~\cite{Takechi}.
}
 \label{Fig:DWS-cross-section}
\end{center}
\end{figure}

Of course, the well-known indicator of nuclear deformation
is the rotational spectra and the $E2$ transition probabilities.
They are, however, difficult to measure in unstable nuclei.
As already discussed in the previous subsection,
the reaction cross section $\sigma_{\rm R}$ can be utilized instead.
Figure \ref{Fig:DWS-cross-section} shows $\sigma_{\rm R}$
for Ne isotopes calculated with the DWS model.
The deformation parameters and pairing gaps are calculated either
with $\widetilde\Delta=12/\sqrt{A}$ or $4/\sqrt{A}$~MeV,
and listed in Table~\ref{tab:WSS_P12-beta} or~\ref{tab:WSS_P04-beta}.
A result of the DWS model
with the deformations calculated with the deformed Gogny-D1S HFB
method~\cite{HFBwoAMP,HFBchart} (Table~\ref{tab:D1SHFB-beta})
is also included as a dotted line.
Comparing with the deformation parameters obtained with the AMD
(Table~\ref{tab:AMD-result}),
the used value of deformation parameters are zero or small around $^{30}$Ne
since the neutron number $N=20$ is a spherical magic number.
As a consequence of this, the reaction cross sections based on
both the microscopic-macroscopic (Strutinsky) and
the deformed Gogny-D1S HFB methods underestimate
the experimental data for $^{28-31}$Ne.
Note that the effect of finite pairing gap alone is small if deformation
is fixed as it is shown in the next subsection
(see Fig.~\ref{Fig:DWS-Pairing}).

The reason why the AMD calculation gives large deformations
in the IOI region is that the optimum deformation is searched
after the angular momentum projection (note that the same Gogny D1S force
is used in the HFB calculation of Refs.~\cite{HFBwoAMP,HFBchart}).
It is known that the potential energy surface as a function of
quadrupole deformation is rather shallow for nuclei in the IOI region.
In such a case, the energy gain of the AMP at large deformation can
easily change the equilibrium deformation, see e.g. Ref.~\cite{RER03}
for Ne isotopes.  The angular momentum projection is thus important
for the IOI region to obtain large deformations.

\subsection{Woods-Saxon model with AMD deformation}
\label{Sec:Epair}

In the previous subsection, it has been shown that the microscopic-macroscopic
method with the DWS model as well as the deformed Gogny-D1S HFB approach
(without the AMP) do not give expected large deformations in the IOI region.
Therefore, we employ, throughout in this subsection,
the deformations obtained by the AMD in the DWS model
and compare the results with the AMD calculations and experimental data.
In most of this subsection, the pairing correlation is neglected;
its effect is discussed at the end.

\begin{table}
\caption{
 Deformation parameter $\beta_2$ and $\gamma$
 deduced from the AMD intrinsic density.
 Those with higher multipoles $\lambda > 2$ are not included. The
 Nilsson asymptotic quantum numbers of last neutron are included as 
 the last column for axially symmetric cases.}
\label{tab:DWS-parameter}
\begin{center}
\begin{tabular}{cccccc}\hline \hline
 nuclide   & $\bar{\beta}$ & $\bar{\gamma}$ &  $\beta_2$ & $\gamma$
 & [$N$,$n_3$,$\Lambda$,$\Omega$] for last-n \\ \hline
 $^{20}$Ne & 0.46 &  0$^{^\circ}$  & 0.479  &  0$^{^\circ}$
 & [2,2,0,1/2]  \\
 $^{21}$Ne & 0.44 &  0$^{^\circ}$  & 0.456  &  0$^{^\circ}$
 & [2,1,1,3/2]  \\
 $^{22}$Ne & 0.39 &  0$^{^\circ}$  & 0.400  &  0$^{^\circ}$
 & [2,1,1,3/2]  \\
 $^{23}$Ne & 0.32 &  0$^{^\circ}$  & 0.325  &  0$^{^\circ}$
 & [2,0,2,5/2]  \\
 $^{24}$Ne & 0.25 & 60$^{^\circ}$  & 0.258  & 60$^{^\circ}$
 & [2,0,0,1/2]  \\
 $^{25}$Ne & 0.20 & 31$^{^\circ}$  & 0.202  & 31.5$^{\circ}$
 &              \\
 $^{26}$Ne & 0.22 & 0.1$^{^\circ}$ & 0.221  &  0$^{^\circ}$
 & [2,1,1,1/2]  \\
 $^{27}$Ne & 0.27 & 13.6$^{^\circ}$& 0.273  & 14.1$^{\circ}$
 &              \\
 $^{28}$Ne & 0.50 &  0$^{^\circ}$  & 0.526  &  0$^{^\circ}$
 & [3,3,0,1/2]  \\
           & 0.28 & 60$^{^\circ}$  & 0.291  & 60$^{^\circ}$
 & [2,1,1,3/2]  \\
 $^{29}$Ne & 0.43 &  0$^{^\circ}$  & 0.445  &  0$^{^\circ}$
 & [2,0,0,1/2]  \\
 $^{30}$Ne & 0.39 &  0$^{^\circ}$  & 0.400  &  0$^{^\circ}$
 & [2,0,0,1/2]  \\
 $^{31}$Ne & 0.41 &  0$^{^\circ}$  & 0.422  &  0$^{^\circ}$
 & [3,2,1,3/2]  \\
 $^{32}$Ne & 0.33 &  0$^{^\circ}$  & 0.335  &  0$^{^\circ}$
 & [2,0,2,3/2]  \\
 \hline
\end{tabular}
\end{center}
\end{table}

\begin{figure}[htbp]
\begin{center}
 \includegraphics[width=0.35\textwidth,clip]{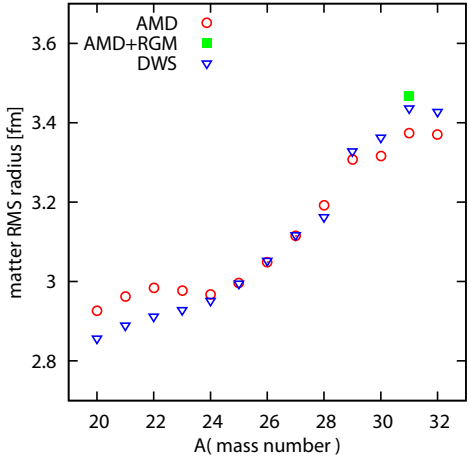}
 \caption{(Color online) Matter RMS radii of Ne isotopes calculated with
 the DWS, AMD and AMD+RGM models.
 The opened circles represent results of the AMD model, and the closed square
 denotes a result of the AMD+RGM model for $^{31}$Ne.
 The opened inverted triangles are results of the deformed Woods-Saxon
 calculation.
 }
 \label{Fig-AMD-DWS-RMS}
\end{center}
\end{figure}

Table \ref{tab:DWS-parameter} lists up the deformation parameters
$\beta_2$ and $\gamma$ deduced from the corresponding AMD values
$\bar{\beta}$ and $\bar{\gamma}$ and used in the following DWS calculations.
In Fig.~\ref{Fig-AMD-DWS-RMS}, the matter RMS radius calculated
with the DWS model is compared with those of the AMD and AMD+RGM calculations.
The DWS model well reproduces the matter RMS radii of the AMD calculation for
$^{24-29}$Ne in which $S_{\rm -1n}$ is large.
For $^{30-32}$Ne in which $S_{\rm -1n}$ is small,
the matter RMS radii of the AMD calculation are slightly
smaller than those of the DWS model. The deviation may come from the fact
that the AMD density is inaccurate in its tail region,
since the DWS model almost
reproduces the matter RMS radius of the AMD+RGM calculation for $^{31}$Ne.
For $^{20-23}$Ne, the matter RMS radii of the AMD calculation are larger
than those of the DWS model. This may imply that the $\alpha$ clustering
is well developed in the AMD calculation and the different type of
deformation from those included in the present Woods-Saxon model,
e.g., the octupole deformation ($\alpha_{3\mu}$), may be important
for $^{20-23}$Ne.

\begin{figure}[htbp]
\begin{center}
\hspace*{-0.05cm}
 \includegraphics[width=0.4\textwidth,clip]{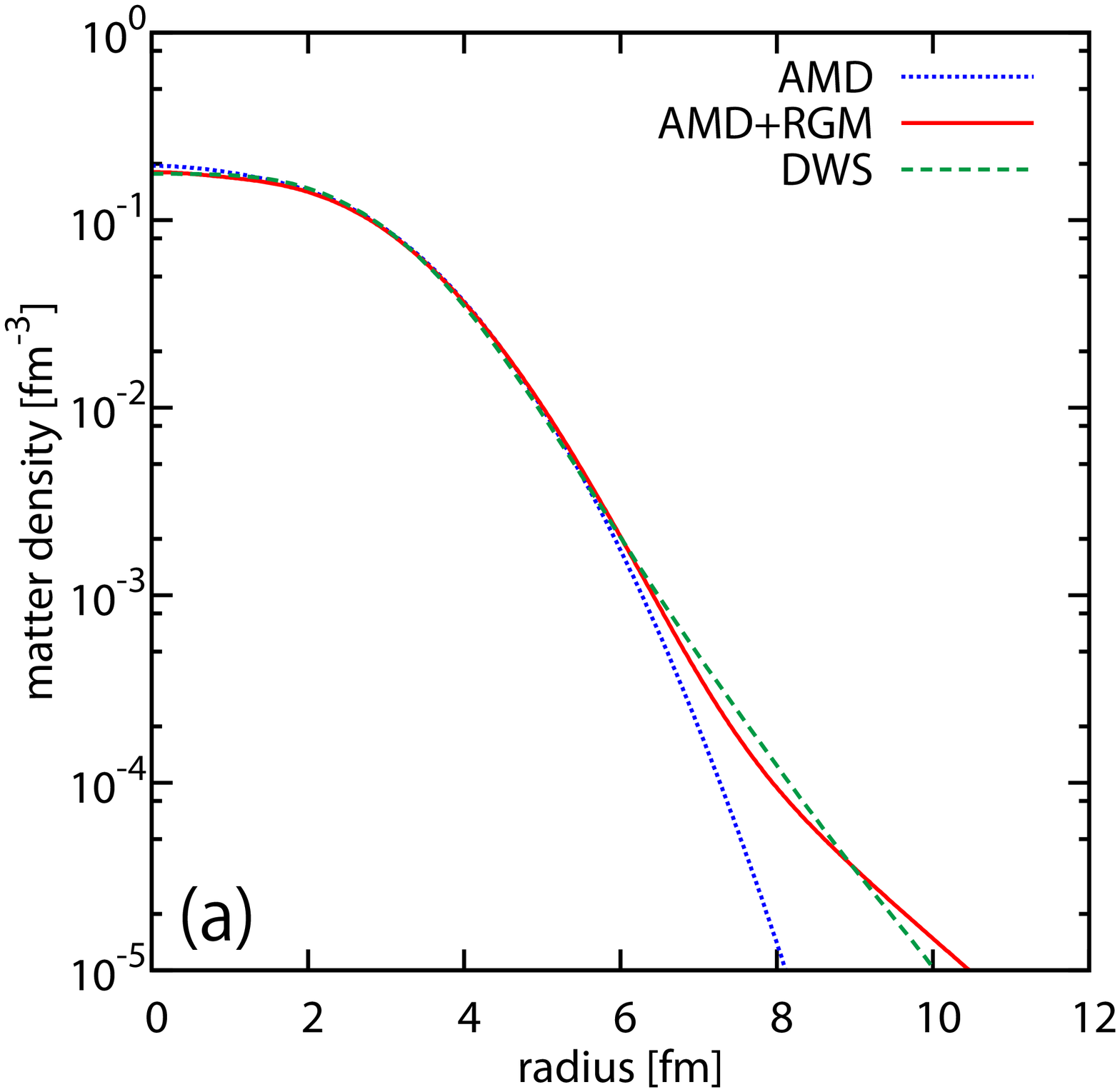}
 \includegraphics[width=0.405\textwidth,clip]{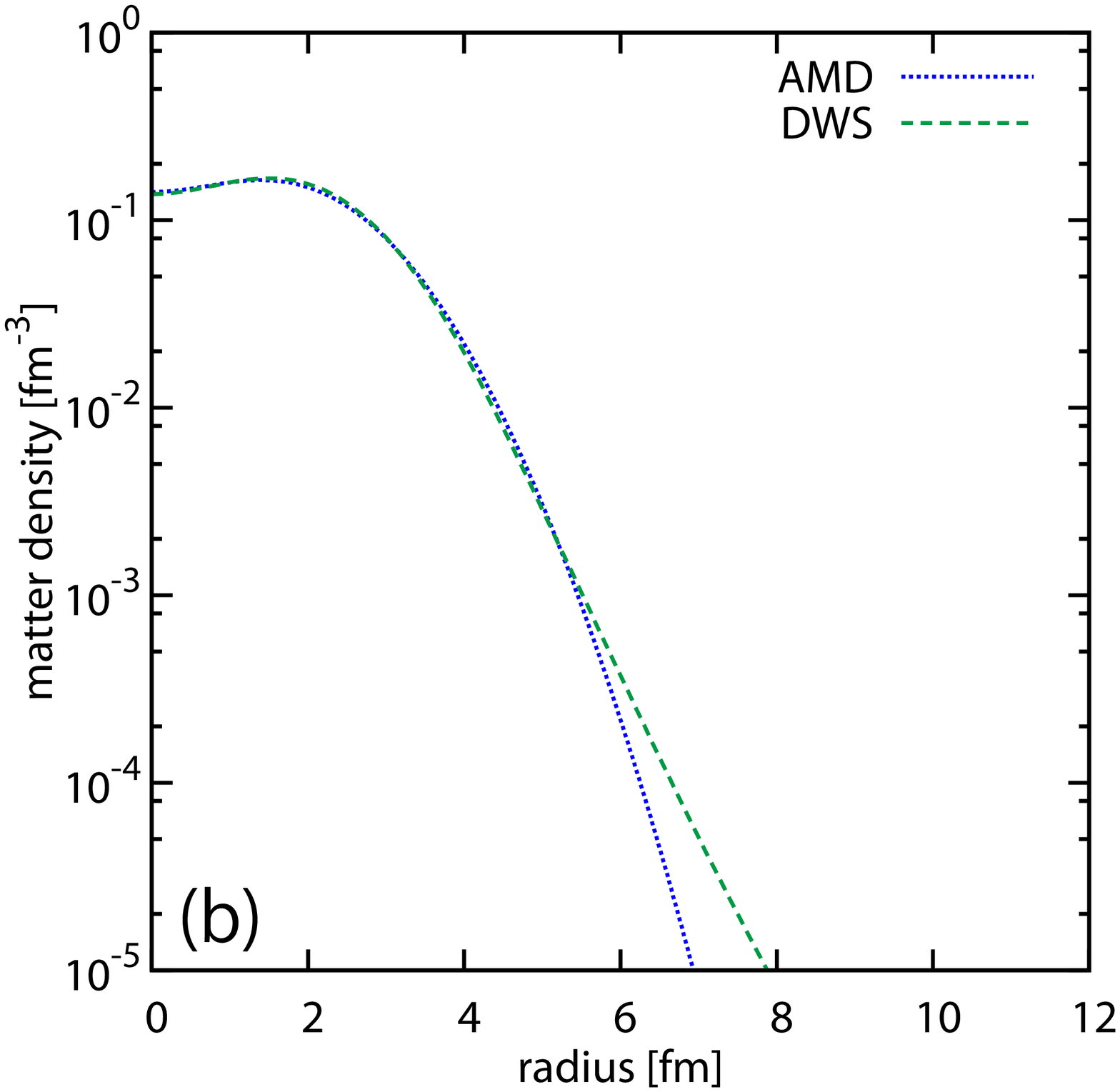}
 \caption{(Color online)
Density distributions of (a) $^{31}$Ne and (b) $^{24}$Ne. 
The dotted line represents a result of the AMD model, whereas 
the dashed line corresponds to a result of the DWS model. 
The solid line is a result of the AMD+RGM model. 
 }
 \label{Fig-density-31-24}
\end{center}
\end{figure}

The nucleon density distributions are plotted in Fig. \ref{Fig-density-31-24}
for $^{24}$Ne and $^{31}$Ne.
The AMD densities (dotted curves) decrease
with increasing $r$ more rapidly than
the densities (dashed curves) of the DWS model.
The deviation between the two densities at large $r$ is rather small
for $^{24}$Ne where $S_{\rm -1n}$ is large.
The deviation is, however,
enlarged for $^{31}$Ne in which $S_{\rm -1n}$ is small.
The AMD density is thus inaccurate at large $r$ particularly for $^{31}$Ne.
A tail-correction to the AMD density can be made by the AMD+RGM calculation.
The density (solid curve) has actually a long-range tail
and consequently becomes close to that of the DWS model.
As an important result, the density of the DWS model almost
agrees with that of the AMD+RGM calculation
for $^{31}$Ne with $\beta_2=0.422$.
This result indicates that the DWS model with the AMD deformation
is considered to be a handy way of making a tail correction to the AMD density.

\begin{figure}[htbp]
\begin{center}
\hspace*{-0.5cm}
 \includegraphics[width=0.35\textwidth,clip]{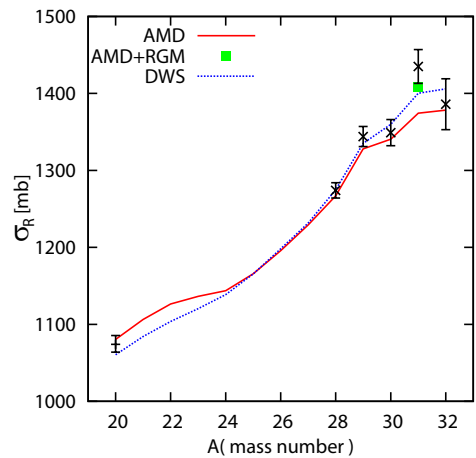}
 \caption{(Color online)
Reaction cross sections for Ne isotopes
calculated with the DWS, AMD and AMD+RGM models.
The dotted line represents results of the DWS model, while
the solid line corresponds to the AMD results. The closed square represents
a result of the AMD+RGM calculation without breakup contribution.
The experimental data are taken from Ref.~\cite{Takechi}.
 }
 \label{Fig-HF-DWSx}
\end{center}
\end{figure}

In Fig.~\ref{Fig-HF-DWSx}, the reaction cross sections
are calculated for Ne isotopes with the DWS, AMD and AMD+RGM models.
The differences among the three calculations are similar to
the corresponding differences for the matter RMS radii shown in
Fig.~\ref{Fig-AMD-DWS-RMS}, as expected.
Note that the result of the DWS model nearly agrees with
that of the AMD+RGM model.
As an important result, the reaction cross sections calculated
with the DWS model are consistent with the experimental data~\cite{Takechi}.

\begin{figure}[htbp]
\begin{center}
 \includegraphics[width=0.355\textwidth,clip]{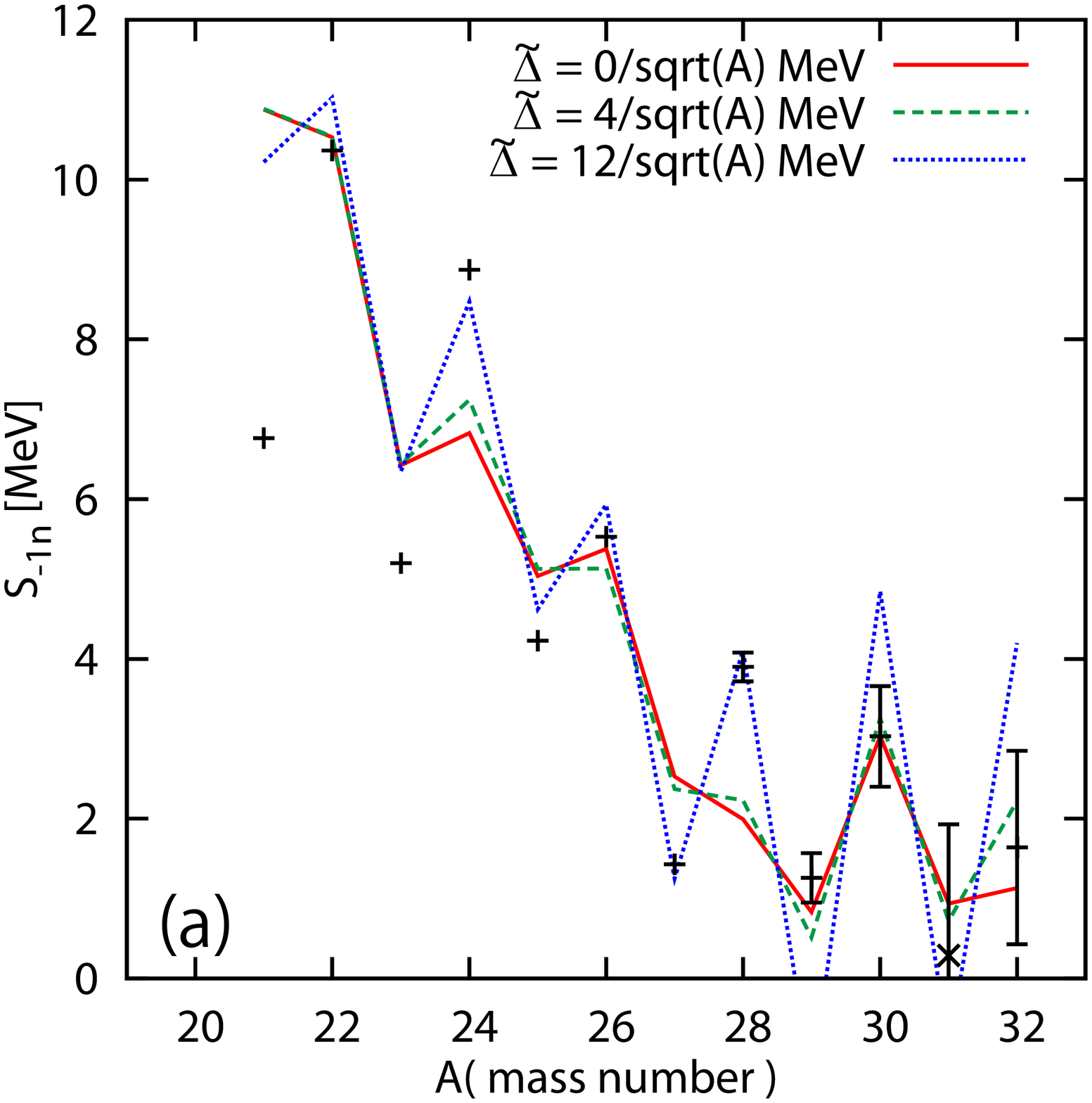}
\hspace*{-0.49cm}
 \includegraphics[width=0.379\textwidth,clip]{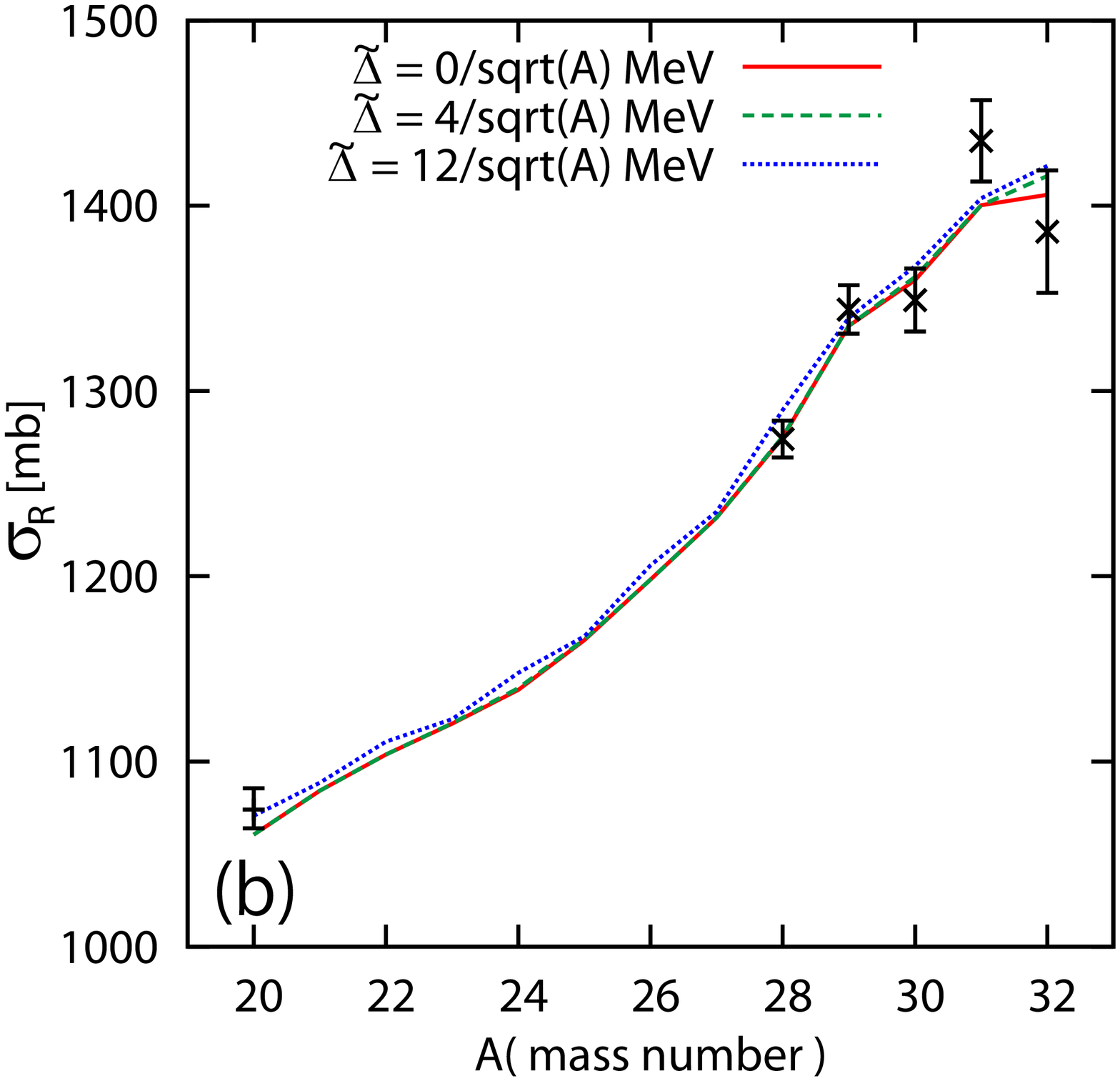}
  \caption{(Color online)
  The pairing effect on (a) the one-neutron separation energy $S_{\rm -1n}$
  and (b) the reaction cross sections $\sigma_{\rm R}$ for Ne isotopes.
  The solid line is a result of the DWS model without pairing correction and
  the dashed (dotted) lines stand for a result
  of the DWS model including the pairing correlation
  with $\widetilde\Delta=4/\sqrt{A}$ $(12/\sqrt{A})$~MeV.
  The experimental data are taken from Ref.~\cite{AusiWapstra,Jurado}
  in panel (a) and from Ref.~\cite{Takechi} in panel (b).
 }
 \label{Fig:DWS-Pairing}
\end{center}
\end{figure}

Finally, the effect of pairing correlation is investigated
for the deformations given by the AMD model.  Two cases of
the pairing strengths given by the standard value $\widetilde\Delta=12/\sqrt{A}$ MeV
and the weaker value $\widetilde\Delta=4/\sqrt{A}$~MeV are considered as
in the previous subsection,
and the result is shown in Fig.~\ref{Fig:DWS-Pairing}.
With the AMD deformations the even-odd effect on the separation energies
$S_{\rm -1n}$ for $^{29-32}$Ne is too much enhanced if the standard
pairing strength is used, while a better fitting is obtained
with the weaker pairing strength, as is shown in Fig.~\ref{Fig:DWS-Pairing}(a).
In Fig.~\ref{Fig:DWS-Pairing}(b), the reaction cross sections
for Ne isotopes are evaluated by the DWS calculations with different
pairing strengths.
The reaction cross sections are enhanced a bit by the pairing effect,
but its effect is small even with the standard pairing strength.

For weakly bound systems, it is speculated that
the pairing correlation leads to an extra binding of halo orbit
and makes the nuclear radius shrink; it is called
the ``pairing anti-halo'' effect~\cite{BDP00,Hagino:2011ji}.
Our Kruppa-BCS method can produce the anti-halo like effect~\cite{OST10},
but a reduction due to the pairing effect
is not observed in the present calculations.
Possible reasons may be the large deformations,
which tend to prevent the anti-halo effect,
and that the binding of the last neutron orbit is not weak enough. 
As a future work, the deformed HFB calculation is highly expected
to answer whether the ``pairing anti-halo'' effect really occurs
and reflects the reaction cross section.

In the DWS model that corresponds to the AMD model with the tail correction, 
$\sigma_{\rm R}$ for $^{32}$Ne is slightly larger than that for $^{31}$Ne 
as shown in Fig.~\ref{Fig-HF-DWSx}, but 
the RMS radius for $^{32}$Ne is smaller  than that for $^{31}$Ne 
as presented in Fig.~\ref{Fig-AMD-DWS-RMS}. 
The reduction of the RMS radius comes from that of $\beta_2$. 
It is interesting to consider what causes the reduction of $\beta_2$. 
This is another interesting future subject related 
to the ``pairing anti-halo'' effect mentioned above.

\section{Summary}
\label{Summary}

We determined deformations of $^{20-32}$Ne with
the fully-microscopic AMD model that has no adjustable parameter.
The quadrupole deformation parameter determined is around 0.4
in the IOI region and $^{31}$Ne is then a halo nuclei
with large deformation.

We have also performed the microscopic-macroscopic (Strutinsky)
calculations with the Woods-Saxon potential, and found that the obtained
deformations are too small, which is consistent with the deformed HFB
calculations without the AMP in this region.

As a reaction model, we used the DFM
with the Melbourne $g$-matrix. 
The microscopic reaction model yields good agreement 
with the measured $\sigma_{\rm R}$ for $^{20,28-32}$Ne, 
if the projectile density is constructed either (I) by the AMP-AMD calculation
with the Gogny D1S interaction 
or (II) by the Woods-Saxon mean-field model
with the deformation obtained by the AMP-AMD calculation.
Method I has no adjustable parameter, but the tail of the density is
inaccurate. We then made a tail correction to the AMD density for $^{31}$Ne 
by using the AMD+RGM method. 
Method II provides the nucleon density with the proper asymptotic form,
but a parameter set of the Woods-Saxon potential should be
carefully chosen.  The parameter set recently proposed 
by R.~Wyss~\cite{WyssPriv} is shown to be very successful:
Method II yields almost the same $\sigma_{\rm R}$
as the AMD+RGM method. This means that Method II is a handy way of
simulating results of Method I with the tail correction.

The two types of DFM well reproduce the measured $\sigma_{\rm R}$ 
for  $^{20, 28-32}$Ne. Deformations of
$^{28-32}$Ne were definitely determined through this analysis. 
This analysis also yields a reasonable prediction for deformations of 
$^{21-27}$Ne. We also showed that the AMP is essential to obtain
the large deformations required for reproducing the measured 
$\sigma_{\rm R}$ for $^{28-32}$Ne but 
that the effect of the BCS-type pairing correlation is small there.

\vspace*{5mm}

\section*{Acknowledgements}

The authors thank M. Takechi for providing the numerical
data and H. Sakurai, M. Fukuda, and D. Baye for useful discussions.
This work is supported in part by Grant-in-Aid for Scientific Research~(C)
No.~22540285 and 22740169 from Japan Society for the Promotion of Science.
The numerical calculations of this work were performed
on the computing system in Research Institute
for Information Technology of Kyushu University.



\begin{thebibliography}{00}

\bibitem{Warburton}
  E.~K.~Warburton, J.~A.~Becker, and B.~A.~Brown,
    Phys.\ Rev.\ C {\bf 41}, 1147 (1990).

\bibitem{Mot95}
T. Motobayashi {\it et al.},
Phys.\ Lett.\ B {\bf 346}, 9 (1995).

\bibitem{Caurier}
  E.~Caurier, F. Nowacki, A. Poves, J. Retamosa,
  Phys.\ Rev.\ C {\bf 58}, 2033 (1998).

\bibitem{Utsuno}
  Y.~Utsuno, T. Otsuka, T. Mizusaki, M. Honma,
  Phys.\ Rev.\ C {\bf 60}, 054315 (1999).

\bibitem{Iwas01}
H. Iwasaki {\it et al.},
Phys.\ Lett.\ B {\bf 522}, 227 (2001).

\bibitem{Yana03}
Y. Yanagisawa {\it et al.},
Phys.\ Lett.\ B {\bf 566}, 84 (2003).


\bibitem{Tanihata}
  I.~Tanihata, \textit{et al}.,
  \newblock
  Phys.\ Lett.\ B {\bf 289}, 261 (1992). \\
  I.~Tanihata, \newblock
   J.\ Phys.\ G {\bf 22}, 157 (1996).

\bibitem{Jensen}
  A.~S.~Jensen, \textit{et al}., \newblock
    Rev.\ Mod.\ Phys.\ {\bf 76}, 215 (2004).

\bibitem{Jonson}
   B. Jonson, \newblock
   Phys.\ Rep.\ {\bf 389}, 1 (2004).


\bibitem{Takechi} M. Takechi {\it et al.}, Nucl. Phys. {\bf A834}, 412c (2010).



\bibitem{Nakamura}
    T.~Nakamura, \textit{et al}.,
    Phys. Rev. Lett. {\bf 103}, 262501 (2009).



\bibitem{Gade}
  A.~Gade, \textit{et al.},
  Phys.\ Rev.\ C {\bf 77}, 044306 (2008).



\bibitem{M3Y}
G. Bertsch, J. Borysowicz, M. McManus, and W.G. Love,
Nucl. Phys. A{\bf 284}, 399(1977).

\bibitem{JLM}
J.-P. Jeukenne, A. Lejeune and C. Mahaux, Phys. Rev. C{\bf 16}, 80 (1977);
ibid. Phys. Rep. {\bf 25}, 83 (1976).


\bibitem{Brieva-Rook}
F.A. Brieva and J.R. Rook, Nucl. Phys. A{\bf 291}, 299 (1977);
ibid. 291, 317 (1977); ibid. 297, 206 (1978).

\bibitem{Satchler-1979}
G. R. Satchler, Phys. Rep. {\bf 55}, 183-254 (1979).

\bibitem{Satchler}
G. R. Satchler, "Direct Nuclear Reactions",
Oxfrod University Press, (1983). 

\bibitem{CEG}
N. Yamaguchi, S. Nagata and T. Matsuda, Prog. Theor.
Phys. {\bf 70}, 459 (1983);
N. Yamaguchi, S. Nagata and J. Michiyama,
Prog. Theor. Phys. {\bf 76}, 1289 (1986).

\bibitem{Rikus-von Geramb}
L. Rikus, K. Nakano and H. V. von Geramb, Nucl. Phys. A{\bf 414}, 413 (1984);
L. Rikus and H.V. von Geramb, Nucl. Phys. A{\bf 426}, 496 (1984).

\bibitem{Amos}
K. Amos, P. J. Dortmans, H. V. von Geramb, S. Karataglidis,
and J. Raynal, in \textit{Advances in Nuclear Physics}, edited by
J. W. Negele and E. Vogt(Plenum, New York, 2000) Vol. 25, p. 275. 

\bibitem{CEG07}
T. Furumoto, Y. Sakuragi, and Y. Yamamoto, Phys. Rev. C{\bf 78},
044610 (2008); {\it ibid.}, C{\bf 79}, 011601(R) (2009);
{\it ibid.}, C{\bf 80}, 044614 (2009).


\bibitem{ERT}
M. Yahiro, K. Ogata, and K. Minomo,
Prog. Theor. Phys. {\bf 126}, 167(2011).

\bibitem{TFH97}
J. Terasaki, H. Flocard, P.-H. Heenen, and P. Bonche,
Nucl.\ Phys.\ A {\bf 621}, 706 (1997).

\bibitem{RER02}
R.~Rodr\'iguez-Guzm\'an, J.L. Egido, and L.M. Robledo,
Nucl.\ Phys.\ A {\bf 709}, 201 (2002).

\bibitem{YG04}
M. Yamagami and Nguyen Van Giai,
Phys.\ Rev.\ C {\bf 69}, 034301 (2004).

\bibitem{RER03}
R.R.~Rodr\'iguez-Guzm\'an, J.L. Egido, and L.M. Robledo,
Eur.\ Phys.\ J.\ A {\bf 17}, 37 (2003).

\bibitem{Kimura}
M. Kimura and H. Horiuchi, Prog. Theor. Phys. 111, 841 (2004).

\bibitem{Kimura1}
M. Kimura, Phys. Rev. C{\bf 75}, 041302 (2007).

\bibitem{Minomo-DWS}
K. Minomo, T. Sumi, M. Kimura, K. Ogata, Y. R. Shimizu, and M. Yahiro,
Phys. Rev. C {\bf 84}, 034602 (2011).


\bibitem{Minomo:2011bb}
  K.~Minomo, T.~Sumi, M.~Kimura, K.~Ogata, Y.~R.~Shimizu and M.~Yahiro,
  arXiv:1110.3867 [nucl-th].





\bibitem{WyssPriv}
R. Wyss, private communication(2005).

\bibitem{Watson}
K. M. Watson, Phys. Rev. {\bf 89}, 115 (1953).

\bibitem{KMT}
A. K. Kerman, H. McManus, and A. M. Thaler, Ann.
Phys.(N.Y.) 8, {\bf 51} (1959).

\bibitem{Yahiro-Glauber}
M.~Yahiro, K.~Minomo, K.~Ogata, and M.~Kawai,
Prog.\ Theor.\ Phys.\ {\bf 120}, 767 (2008).


\bibitem{DFM-standard-form}
B. Sinha, Phys. Rep. {\bf 20}, 1 (1975). \\
B. Sinha and S. A. Moszkowski, Phys. Lett. B{\bf 81}, 289 (1979).

\bibitem{DFM-standard-form-2}
T. Furumoto, Y. Sakuragi, and Y. Yamamoto, Phys. Rev. C{\bf 82},
044612 (2010).

\bibitem{Minomo:2009ds}
  K.~Minomo, K.~Ogata, M.~Kohno, Y.~R.~Shimizu and M.~Yahiro,
  J.\ Phys.\ G {\bf 37}, 085011 (2010)
  [arXiv:0911.1184 [nucl-th]].


\bibitem{BonnB}
R. Machleidt, K. Holinde K and Ch. Elster,
Phys. Rep. {\bf 149}, 1(1987).







\bibitem{GognyD1S}
J.~F.~Berger, M.~Girod and D.~Gogny,
Comput. Phys. Commun. {\bf 63}, 365 (1991).

\bibitem{NS81}
W.~Nazarewicz and A.~Sobiczewski, Nucl. Phys. {\bf A369}, 396 (1981).

\bibitem{SS09}
T. Shoji and Y. R. Shimizu,
Progr.\ Theor.\ Phys.\ {\bf 121}, 319 (2009).

\bibitem{CDN87}
S.~Cwiok, J.~Dudek, W.~Nazarewicz, J.~Skalski and T.~Werner,
Comp.\ Phys.\ Comm.\ {\bf 46}, 379 (1987).

\bibitem{Str68}
M.~Strutinsky, Nucl. Phys. {\bf A122}, 1 (1968).

\bibitem{FunnyHills}
M.~Brack, J.~Damg{\aa}rd, A.~S.~Jensen, H.~C.~Pauli, V.~M.~Strutinsky,
and C.~Y.~Wong, Rev. Mod. Phys. {\bf 44}, 320 (1972).

\bibitem{MS67}
W.~D.~Myers and W.~J.~Swiatecki,
Nucl.\ Phys.\ {\bf 81} 1 (1966);
Ark. Phys. {\bf 36}, 343 (1967).

\bibitem{TST10}
N.~Tajima, Y.~R.~Shimizu, and S.~Takahara,
Phys. Rev. C {\bf 82} (2010), 034316.

\bibitem{RS80}
P.~Ring and P.~Schuck,
{\em The nuclear many-body problem}, Springer, New York (1980).


\bibitem{OST10}
T.~Ono, Y.~R.~Shimizu, N.~Tajima and S.~Takahara,
Phys.\ Rev.\ C {\bf 82}, 034310 (2010).


\bibitem{TStobeP}
S.~Tagami and Y.~R.~Shimizu, to be published.

\bibitem{TB-1958}
L.J. Tassie and F.C. Barker,
Phys. Rev. {\bf 111}, 940(1958).

\bibitem{Suzuki-C75}
W. Horiuchi, Y. Suzuki, B. Abu-Ibrahim and A. Kohama,
Phys. Rev. C {\bf 75}, 044607(2007).

\bibitem{expC12C12}
 M. Takechi, \textit{et al}.,
  Phys.\ Rev.\ C \textbf{79}, 061601(R) (2009).

\bibitem{Ne20-sigmaI}
L. Chulkov \textit{et al}.,
Nucl. Phys. \textbf{A603} 219, (1996).

\bibitem{Na23-sigmaI}
T. Suzuki \textit{et al}.,
Phys. Rev. Lett. \textbf{75}, 3241 (1995).

\bibitem{C12-density}
H. de Vries, C. W. de Jager, and C. de Vries,
At. Data Nucl. Data Tables \textbf{36}, 495 (1987).

\bibitem{Singhal}
R. P. Singhal \textit{et al}., Nucl. Instr. and Meth. 148, 113(1978).


\bibitem{Love-Franey}
W. G. Love and M. A. Franey,
Phys. Rev. C{\bf 24}, 1073 (1981);
M. A. Franey and W. G. Love,
Phys. Rev. C{\bf 31}, 488 (1985).

\bibitem{Paris}
M. Lacombe,
 \textit{et al}.,
 Phys. Rev. C\textbf{21}, 861(1980).


\bibitem{ICN} http://www.nnbc.bnl.gov/chart/


\bibitem{Jurado}
B. Jurado, \textit{et al}.,
 Phys. Lett. \textbf{B649}, 43 (2007).

\bibitem{AusiWapstra}
G. Ausi,
 A. H. Wapstra, and C. Thibault,
 Nucl. Phys. \textbf{A729}, 337, (2003).



\bibitem{CDCC-review1}
M.~Kamimura, 
M.~Yahiro, Y.~Iseri, Y.~Sakuragi, H.~Kameyama, and
M.~Kawai, \newblock
Prog.\ Theor.\ Phys.\ Suppl.\ {\bf 89}, 1 (1986).

\bibitem{CDCC-review2}
N.~Austern, 
Y.~Iseri, M.~Kamimura, M.~Kawai, G.~Rawitscher, and
M.~Yahiro, \newblock
Phys.\ Rep.\ {\bf 154}, 125 (1987).

\bibitem{CDCC-foundation1}
N.~Austern, M.~Yahiro, and M.~Kawai,
\newblock
Phys.\ Rev.\ Lett. {\bf 63}, 2649 (1989).

\bibitem{CDCC-foundation2}
N.~Austern, 
M.~Kawai, and  M.~Yahiro, \newblock
Phys.\ Rev.\ C {\bf 53}, 314 (1996).

\bibitem{CDCC-foundation3}
A. Deltuva, A.M.~Moro, E.~Cravo, F.M.~Nunes, and A.C.~Fonseca,
Phys.\ Rev.\ C {\bf 76}, 064602 (2007).

\bibitem{Rusek}
K. Rusek and K.W. Kemper,
Phys. Rev. C {\bf 61}, 034608 (2000).

\bibitem{Tostevin2}
J.A. Tostevin, F.M. Nunes, and I.J. Thompson,
Phys. Rev. C {\bf 63}, 024617 (2001).

\bibitem{Davids}
B. Davids, S.M. Austin, D. Bazin, H. Esbensen,
B.M. Sherrill, I.J. Thompson, and J.A. Tostevin,
Phys. Rev. C {\bf 63}, 065806 (2001).

\bibitem{Mortimer}
J. Mortimer, I. J. Thompson, and J. A. Tostevin,
Phys.\ Rev.\ C {\bf 65}, 064619 (2002).

\bibitem{Eikonal-CDCC}
K.~Ogata, M.~Yahiro, Y.~Iseri, T.~Matsumoto, and M.~Kamimura,
Phys.\ Rev.\ C {\bf 68}, 064609 (2003).


\bibitem{Matsumoto3}
T.~Matsumoto, 
E.~Hiyama, K.~Ogata, Y.~Iseri, M.~Kamimura,
S.~Chiba, and M.~Yahiro,
Phys.\ Rev.\ C {\bf 70}, 061601(R) (2004).

\bibitem{Howell}
D.J. Howell, J.A. Tostevin, and J.S. Al-Khalili,
J. Phys. G: Nucl. Part. Phys. {\bf 31}, S1881 (2005).

\bibitem{Rusek2}
K. Rusek, I. Martel, J. G\'{o}mez-Camacho,
A.M. Moro, and R. Raabe,
Phys. Rev. C {\bf 72}, 037603 (2005).

\bibitem{Matsumoto4}
T.~Matsumoto, 
T.~Egami, K.~Ogata, Y.~Iseri, M.~Kamimura, and M.~Yahiro,
Phys.\ Rev.\ C {\bf 73}, 051602(R) (2006).

\bibitem{Moro}
A.M. Moro, K. Rusek, J.M. Arias, J. G\'{o}mez-Camacho,
and M. Rodr\'{i}guez-Gallardo,
Phys. Rev. C {\bf 75}, 064607 (2007).

\bibitem{THO-CDCC}
M.~Rodr\'{i}guez-Gallardo, 
J.~M.~Arias, J.~G\'{o}mez-Camacho,
R.~C.~Johnson, A.~M.~Moro, I.~J.~Thompson, and J.~A.~Tostevin,
\newblock
Phys.\ Rev.\ C {\bf 77}, 064609 (2008).

\bibitem{4body-CDCC-bin}
M.~Rodr\'{i}guez-Gallardo,
J. M. Arias, J. G\'{o}mez-Camacho,
A. M. Moro, I. J. Thompson, and J. A. Tostevin,
Phys.\ Rev.\ C {\bf 80}, 051601(R) (2009).

\bibitem{Matsumoto:2010mi}
T.~Matsumoto, K.~Kato, and M.~Yahiro,
Phys.\ Rev.\  C {\bf 82}, 051602 (2010)
[arXiv:1006.0668 [nucl-th]].

\bibitem{Avrigeanu}
M. Avrigeanu and A.M. Moro,
Phys. Rev. C {\bf 82}, 037601 (2010).


\bibitem{30Ne}
W. Horiuchi, Y. Suzuki, P. Capel, and D. Baye,
Phys. Rev. C \textbf{81}, 024606 (2010).

\bibitem{SDN98}
W.~Satula, J.~Dobaczewski and W.~Nazarewicz,
Phys.\ Rev.\ Lett.\ {\bf 81}, 3599 (1998).

\bibitem{OMA06}
T.~Otsuka, T.~Matsuo and D.~Abe,
Phys.\ Rev.\ Lett.\ {\bf 97}, 162501 (2006).

\bibitem{HFBwoAMP}
S. Hilaire and M. Girod, Eur. Phys. J. A \textbf{33}, 237(2007).

\bibitem{HFBchart}
http://www-phynu.cea.fr/science\_en\_ligne/carte\_potentiels
\_microscopiques/carte\_potentiel\_nucleaire\_eng.htm


\bibitem{BDP00}
K.~Bennaceur,l J.~Dobaczewski, and M.~Ploszajczak,
Phys.\ Lett.\ {\bf B496}, 154 (2000).

\bibitem{Hagino:2011ji}
  K.~Hagino and H.~Sagawa,
  Phys.\ Rev.\  C {\bf 84}, 011303 (2011)
  [arXiv:1105.5469 [nucl-th]].




\end{thebibliography}
\end{document}